\documentclass[aps,superscriptaddress,eqsecnum,nofootinbib,showpacs,
preprintnumbers]{revtex4}

\usepackage{amsmath}
\usepackage{amsfonts}
\usepackage{amssymb}
\usepackage{mathtools}
\usepackage{bm}
\usepackage{xcolor}
\usepackage{hyperref}
\usepackage{orcidlink}

\usepackage{graphicx}
\usepackage{subcaption}

\usepackage{color}
\usepackage{relsize}

\newcommand{\be}{\begin{equation}}
\newcommand{\ee}{\end{equation}}
\newcommand{\bea}{\begin{eqnarray}}
\newcommand{\eea}{\end{eqnarray}}

\begin{document}

\title{Interacting  dilatonic ghost condensate as  dark energy model}

\author{Manuel Gonzalez-Espinoza\orcidlink{0000-0003-0961-8029}}
\email{manuel.gonzalez@upla.cl}
\affiliation{Laboratorio de Did\'actica de la  F\'isica, Departamento de Matem\'atica\text{,} F\'isica y Computaci\'on, Facultad de Ciencias Naturales y Exactas, Universidad de Playa Ancha, Subida Leopoldo Carvallo 270, Valpara\'iso, Chile}
\affiliation{Laboratorio de investigación de Cómputo de Física, Facultad de Ciencias Naturales y
Exactas, Universidad de Playa Ancha, Subida Leopoldo Carvallo 270, Valparaíso, Chile}

\author{Ram\'on Herrera\orcidlink{0000-0002-6841-1629}}
\email{ramon.herrera@pucv.cl}
\affiliation{Instituto de F\'{\i}sica, Pontificia Universidad Cat\'olica de 
Valpara\'{\i}so, 
Casilla 4950, Valpara\'{\i}so, Chile}

\date{\today}

\begin{abstract} 
In this paper, we examine the cosmological evolution of a dilatonic ghost condensate field responsible for dark energy, which interacts with dark matter through a source term. We explore three different interaction models to describe the present universe. For each interaction model, we perform a detailed phase-space analysis, obtaining the stability conditions, and identifying the critical points. Furthermore, we compare our interaction models with the most recent Hubble parameter and supernova Ia data as functions of redshift. Additionally, we investigate the conditions for scaling regimes in these models and analyze the successful transition toward an attractor point to characterize the behavior of dark matter.
\end{abstract}


\maketitle

\section{Introduction}\label{Introduction}

That our Universe is expanding, first revealed by Type Ia Supernovae in the late 1990s \cite{Riess:1998cb, Perlmutter:1998np}, remains one of the most profound puzzles in modern cosmology. The prevailing approach within the $\Lambda$CDM framework attributes this acceleration to a cosmological constant $\Lambda$, a simple yet theoretically problematic term whose extremely small value introduces significant tension with quantum field theory expectations \cite{Martin:2012bt, Copeland:2006wr, Bull:2015stt}. Beyond the theoretical fine-tuning, the $\Lambda$CDM model also faces growing discrepancies with observations, notably in the values of the Hubble parameter ($H_0$) and the amplitude of matter perturbations ($\sigma_8$), suggesting that our current cosmological paradigm may be incomplete \cite{DiValentino:2020zio, DiValentino:2020vvd, Riess:2019cxk, McGaugh:2023nkc}.

So, physicists have explored a range of dynamical models to complement the cosmological constant idea. Among these, scalar field theories have gained particular attention for their versatility. Quintessence \cite{Wetterich:1987fm, Ratra:1987rm, Carroll:1998zi, Tsujikawa:2013fta} and k-essence models \cite{Chiba:1999ka, ArmendarizPicon:2000dh, ArmendarizPicon:2000ah} are prominent examples in which a scalar degree of freedom evolves dynamically to drive cosmic acceleration. Extensions of these frameworks, including scalar fields with non-minimal couplings to matter or curvature \cite{Perrotta:1999am, Faraoni:2000wk}, as well as Galileon-type theories \cite{Deffayet:2009wt, Nicolis:2008in}, have been widely explored for their ability to address the limitations of $\Lambda$CDM and provide better fits to observational data.

One particularly intriguing class of models is based on ghost condensate scalar fields with a negative kinetic term, which under normal circumstances would signal a pathological instability. Remarkably, this problem can be solved if higher-order kinetic corrections stabilize the vacuum at a non-trivial field configuration \cite{Arkani-Hamed:2003pdi, Arkani-Hamed:2003juy, Copeland:2006wr}. In such models, the scalar field evolves towards a vacuum state characterized not by a constant field value, but by a constant time derivative. This ``ghost condensate" background dynamically breaks Lorentz symmetry and acts as a cosmological fluid that can drive acceleration without the drawbacks of a cosmological constant.

However, a continuous issue in standard ghost condensate scenarios is the lack of inhibition of the stabilizing higher-order terms due to the low energy scale of dark energy (DE) relative to the Planck scale. To address this problem, dilatonic extensions have been proposed, where the higher-order kinetic terms are multiplied by an exponential function of a second scalar field, the dilaton. And, originally inspired by low-energy corrections from string theory \cite{Piazza:2004df}, the dilatonic ghost condensate model ensures that the stabilizing terms stay dynamically important even at late times. This helps maintain vacuum stability while still allowing for cosmic acceleration \cite{Gumjudpai:2005ry, Peirone:2019aua}.

Interestingly, models based on dilatonic ghost condensates often emerge from low-energy effective theories that arise in the string landscape. In many of these scenarios, the mechanisms used to stabilize moduli tend to produce scalar fields with non-standard kinetic terms. Even more, what makes these models especially appealing is that they can be naturally incorporated into multiple theoretical frameworks, such as scalar-tensor theories \cite{DeFelice:2010aj, Horndeski:1974wa} and generalized Galileons \cite{Kobayashi:2011nu}. And this opens up promising possibilities for constructing unified models for cosmic acceleration. Moreover, their dynamical behavior has produced attractor solutions consistent with late-time cosmic acceleration and transitions from matter domination \cite{Tsujikawa:2010sc, Frusciante:2018vht}.

On the other hand, in the standard cosmological model, dark energy and dark matter (DM) evolve independently and interact only gravitationally. Furthermore, since there is no fundamental reason to rule out a non-gravitational coupling between dark matter and dark energy, the idea of an interaction between them has become increasingly significant in the search for consistent dark energy scenarios. Indeed, a dynamical exchange of energy or momentum between DM and DE could provide a natural solution to the so-called cosmic coincidence problem, which questions why the energy densities of DM and DE are of the same order of magnitude today \cite{amendola2010dark, Cid:2020kpp, wang2016dark}.

From a theoretical perspective, interacting DE-DM models can be motivated within quantum field theory in curved spacetime, where couplings between scalar fields and matter arise via radiative corrections or as part of effective actions in infrared-modified theories of gravity \cite{Linde:1982zj, Birrell:1982ix}. Phenomenologically, allowing for a coupling between the dark components leads to a modification of the standard energy-momentum conservation equations via a source term \( Q \), representing the energy transfer between them \cite{Amendola:1999qq}. This idea has been explored in depth and has yielded rich dynamics, including scaling solutions, attractors, and scenarios where the equation-of-state parameter crosses the phantom divide \( w = -1 \) without the need for exotic fields \cite{He:2008si, Valiviita:2008iv}.

But models incorporating interactions mostly rely on phenomenological source terms ($Q$) in the conservation equations, allowing the dark matter and dark energy components to exchange energy during cosmic evolution. Even more, these new models, including interacting terms, have demonstrated rich dynamical behaviors, including scaling solutions and late-time attractors, which can potentially account for observational tensions and provide new insights into the evolution of cosmic structures \cite{Copeland:2006wr, Kadam:2024fgz, Gonzalez-Espinoza:2022hui}.

Recent studies have shown that specific forms of these couplings can alleviate tension in $H_0$ and $\sigma_8$, supporting the idea of exploring new interaction terms in different modified gravity theories \cite{rodriguez2020universe, wang2024further}. Importantly, interactions within the dark sector may also affect the growth of cosmic structures and leave detectable imprints in cosmological observables, such as the matter power spectrum, redshift-space distortions, and the cosmic microwave background \cite{DiValentino:2019jae, Kumar:2017dnp}. Hence, the development of viable interacting DE-DM models represents a key avenue to reconcile theoretical consistency with observational data.

Despite the interest in dilatonic ghost condensate fields and the interest in interacting dark-energy models, the combination of these frameworks remains underexplored. On the one hand, the dilatonic ghost condensate provides a well-motivated mechanism for generating scalar fields with non-trivial dynamics. On the other hand, allowing interactions within the dark sector introduces new degrees of freedom that can shape cosmic evolution and account for observational tensions \cite{Copeland:2006wr, Kadam:2024fgz, Gonzalez-Espinoza:2022hui}. Therefore, integrating these two ideas into a unified framework could yield new cosmological behaviors and observational signatures distinct from those of standard scalar field models.

In this work, we aim to investigate the cosmological implications of a dilatonic ghost condensate scalar field acting as DE, while interacting with DM through a phenomenological source term. In this sense, we study how different interactions that enter the energy balance equations for the dilatonic ghost condensate scalar field and DM affect the evolution of the present Universe. In this form, we consider several forms of the interaction kernel $Q$ that have been proposed in the literature, and analyze the dynamical system derived from the Friedmann equations and the evolution of the scalar field. Then, we identify critical points and their stability, allowing us to classify the possible cosmic histories associated with each interaction model. This theoretical framework is then tested against current observational constraints, including Hubble parameter measurements and Type Ia supernova data, to assess its phenomenological viability.

This article is organized as follows. In the next section, we briefly describe our interaction model within the framework of a dilatonic ghost condensate theory. In Section III, we analyze the dynamical system for three different interactions, identifying the critical points and the stability conditions of our DE-DM models. In Section IV, we examine the numerical results of our autonomous system, comparing them with recent data from the Hubble parameter and Type Ia supernovae. Finally, in Section V, we summarize our findings. 

\section{Interaction in a  dilatonic ghost condensate theory}\label{model}

In order to describe our model, we begin with a general four-dimensional effective action  defined  as \cite{Armendariz-Picon:1999hyi,Garriga:1999vw}
\bea
S&=&\int \Bigg[\frac{R}{2\kappa^2} +\mathcal{L}(X,\phi)
\Bigg]\,\sqrt{-g}\,d^{4}{x} 
\, + S_{m}+ S_{r},\label{action}
\eea 
where $R$ corresponds to the scalar Ricci, the quantity $\mathcal{L}(X,\phi)$ denotes the Lagrangian density associated to the kinetic term 
$X= -\frac{1}{2} \partial_{\mu}{\phi}\partial^{\mu}{\phi}$  and to the scalar field $\phi$.
The quantity $g$ is the determinant of the metric $g_{\mu\nu}$, 
and $\kappa^2=8\pi G=M_p^{-2}$, in which the quantity  $M_p$ denotes  the Planck mass. Furthermore, the quantities $S_m$ and $S_r$ correspond to the actions associated with matter and radiation, respectively.

By considering  the effective action given by Eq.(\ref{action}), we can characterize  that the energy density $\rho_{de}$ and the pressure $p_{de}$ related to the DE  
as a function of the scalar field $\phi$ and $X$ assuming a perfect fluid  are defined as\cite{Armendariz-Picon:1999hyi,Garriga:1999vw}
\begin{equation}
\rho_{de}=2X\frac{\partial \mathcal{L}(X,\phi) }{\partial X}-\mathcal{L}(X,\phi),\,\,\,\,\,\,\mbox{and}\,\,\,\,\,\,\,\,\,p_{de}=\mathcal{L}(X,\phi),\label{rp}
\end{equation}
 respectively. Here we note that for the special situation in which $\mathcal{L}(X,\phi)=X-V(\phi)$,   where $V(\phi)$ corresponds to the effective potential associated with the scalar field, the energy density and pressure are reduced to the standard results for the energy density and pressure related to a scalar field in the framework of canonical theory\cite{Chiba:1999ka,Ratra:1987rm}.

In the context  of the  Lagrangian density related to the scalar field, we can consider that this Lagrangian can be written  in the form \cite{Armendariz-Picon:1999hyi}
\begin{equation}
    \mathcal{L}(X,\phi)=\sum_{n\geq 0}\alpha_n(\phi)X^{n+1}-V(\phi),
    \label{exp1}
\end{equation}
where the different parameters  $\alpha_n(\phi)$ are  arbitrary functions of the scalar field $\phi$. For the standard case of canonical theory, we need to consider the values $n=0$ and $\alpha_0(\phi)=1$, respectively.

In the following, we will assume the specific quantities of $n=0$ and $n=1$, in which the functions $\alpha_n(\phi)$ are given by $\alpha_0(\phi)=\alpha$, a constant and $\alpha_1(\phi)=F(\phi)=F$, a function exclusively dependent on the scalar field.   Thus,  the Lagrangian density related to the inflaton field can be written as \cite{Armendariz-Picon:1999hyi} 
\begin{equation}
\label{p}
   \mathcal{L}(X,\phi)=K(X,\phi)-V(\phi)=\alpha\,X+F(\phi)X^2-V(\phi),
\end{equation}
where the term $K(X,\phi)=\alpha \,X+F(\phi)X^2$ denotes  an arbitrary function related with a non-linear kinetic term. In addition, we assume that the quantity $\alpha$   is a dimensionless constant and that
the function $F(\phi)$ has units of $\kappa^2=M_p^{-4}$. 

The motivation for considering the Lagrangian density  given  by Eq.(\ref{p}) emerges from  string-loop corrections, which induce a non-trivial moduli field dependence in the different coefficients associated with the different kinetic terms. Here,
the kinetic terms emerge due to the massive modes of the string in the low-energy action; for more details, see Refs.\cite{Damour:1994zq,Damour:1995pd,Foffa:1999dv}. Thus, motivated by these low-energy effective actions, we shall assume the simplest model to study the interaction between dark energy, associated to the scalar field in which the Lagrangian density takes the specific form of Eq.(\ref{p}) and dark matter characterized by the energy density $\rho_m$.   

To study the cosmological background  dynamics of our model, we consider that the scalar field corresponds to a  homogeneous scalar field i.e., $\phi=\phi(t)$ together with 
a spatially flat Friedmann-Lemaître-Robertson-Walker
(FLRW)  metric  given by
\begin{equation}
ds^2=-dt^2+a^2\,\delta_{ij} dx^i dx^j \,,
\label{FRWMetric}
\end{equation}
where $a=a(t)$ denotes  the scale factor, which   is a function of cosmic time $t$. 

In this way, using Eqs.(\ref{rp}) and (\ref{p}), 
 we determine  that the effective energy density $\rho_{de}$ and pressure $p_{de}$ associated to dark energy  are given by\cite{Armendariz-Picon:1999hyi}  
\bea
\label{rhode}
\rho_{de}&= & \alpha \frac{\dot{\phi }^2}{2}  + \dfrac{3}{4} F \dot{\phi}^4+V,\,\,\,\,\,\,\mbox{and}\,\,\,\,\,\,\,\,
 p_{de}=  \alpha \frac{\dot{\phi }^2}{2} + \dfrac{1}{4} F \dot{\phi}^4-V,
 \eea 
respectively.

In relation to the Friedmann equation, we can write this  equation of motion in terms of the individual energy densities as
\bea
\label{SH00}
&& \frac{3}{\kappa^2} H^2=\rho=\rho_{de}+\rho_{m}+\rho_{r},
\eea
and the equation for $\dot{H}$ as
\bea
 -\frac{2}{\kappa^2} \dot{H}=\rho_{de}+p_{de}+\rho_{m}+\frac{4}{3}\rho_{r},
\label{SHii}
\eea 
where $H=\dot{a}/a$ corresponds to the Hubble parameter and $\rho$ denotes the total energy density. Here, $\rho_m$ and $\rho_r$ correspond to the energy densities associated with matter and radiation, respectively. In addition,
in the following, the dots denote the differentiation with respect to time $t$.

On the other hand, by conserving the total energy-momentum tensor, we can write
\bea
\dot{\rho}+3H(\rho+p)=0.
\eea
In this context, to study the interaction between dark matter and dark energy,
we can assume that these components are coupled through a
source term denoted by $Q$, which enters the energy balance equations for dark matter and dark energy, respectively. Thus, the equations of motion assuming an interaction between the dark sector can be written as follows
\begin{eqnarray}
\dot{\rho}_{de}+3H(\rho_{de}+p_{de})=-Q,\label{ecde}
\end{eqnarray} 
and
\bea
\label{rho_m}
 \dot{\rho}_{m}+3 H\rho_{m}=Q.
\eea
In relation to the interaction term $Q$, in the literature different types of interactions have been studied\cite{Amendola:1999er,Chimento:2003iea,Wang:2016lxa}. This interaction may lead to additional consequences. It could shift the onset of the accelerated expansion era to higher redshifts and might misleadingly indicate a phantom-like equation of state for dark energy \cite{Amendola:2006ku}. Furthermore, the interaction term $Q$ can influence  the result of the fluctuations in the counts of galaxy clusters with redshift, see e.g.\cite{Manera:2005ct}. Similarly, the coupling between the DM and DE can modify the isothermal Maxwell-Boltzmann velocity distribution of weakly interacting massive particles in galaxy halos as was found in Refs.\cite{Tetradis:2005me,Tetradis:2006md}. 

Furthermore, for radiation, the equation of motion for the radiation field yields
\bea
\dot{\rho}_{r}+4 H\rho_{r}=0.
\label{rho_r}
\eea
We note that in view of Eq.(\ref{rhode}), the equation of motion for the dark energy given by Eq.(\ref{ecde}) can be rewritten in terms  of the scalar field as 
\bea
(3 F \dot{\phi}^2+ \alpha)\ddot{\phi }+3 H \dot{\phi }(F \dot{\phi}^2 + \alpha ) 
+V_{,\phi} + \dfrac{3}{4} F_{,\phi} \dot{\phi}^4 =-\dfrac{Q}{\dot{\phi}}.
\label{MFreq2}
\eea 
Besides, we find that the  Friedmann equations can be written as
\bea
\label{MFreq1}
 \frac{3 H^2}{\kappa ^2} &=& \left(V + \alpha  \frac{\dot{\phi }^2}{2}  + \dfrac{3}{4} F \dot{\phi}^4\right)
+\rho _m+\rho _r,
\eea
and the equation for $\dot{H}$ as
\bea
 -\frac{2 \dot{H}}{\kappa ^2} &=&  \alpha \dot{\phi }^2 + F \dot{\phi}^4 
+\rho _m+\frac{4 \rho _r}{3}.\label{ii}
\eea 
Here, we note that the above equation does not depend on the effective potential $V$. In the following, the notation $V_{,\phi}$ denotes $\partial V/\partial \phi$ and  $F_{,\phi}=\partial F/\partial\phi$, respectively.

Additionally, we can define the equation-of-state (EoS) parameter $w_{de}$, associated to the dark energy  as 
\begin{equation}
w_{de}=\frac{p_{de}}{\rho _{de}}=\frac{\left(\alpha \frac{\dot{\phi }^2}{2} + \dfrac{1}{4} F \dot{\phi}^4-V\right)}{\left(\alpha \frac{\dot{\phi }^2}{2}  + \dfrac{3}{4} F \dot{\phi}^4+V\right)},
\label{wDE1}
\end{equation}
and also it is useful to define  the
total EoS parameter $w_{tot}$ as 
\be
w_{tot}=\frac{p}{\rho}=\frac{p_{de}+p_r}{\rho _{de}+\rho _m+\rho _r}.
\label{wtot}
\ee 
In relation to the total EoS parameter, we can associate this parameter  with 
 the deceleration parameter $q$ through
\be
q=-\frac{\ddot{a}a}{\dot{a}^2}=\frac{1}
{2}\left(1+3w_{tot}\right).
\label{deccelparam}
\ee 
Thus, the acceleration of the universe ($\ddot{a}>0$) takes place when the parameter  $q<0$ or equivalently when the total EoS parameter $w_{total}<-1/3$.

Additionally, we  introduce the dimensionless 
 density parameters  associated to DM,DE and the radiation field defined as 
\bea
&& \Omega_{m}\equiv\frac{\kappa^2 \rho_{m}}{3 H^2},\:\:\:\: \Omega_{de}\equiv\frac{\kappa^2 \rho_{de}}{3 H^2},\:\:\:\: \Omega_{r}\equiv \frac{\kappa^2 \rho_{r}}{3 H^2}.
\eea 
In this way, the Friedmann equation given by Eq.(\ref{SH00}) becomes
\be
\Omega_{de}+\Omega_{m}+\Omega_{r}=1.\label{ec}
\ee 
In the following, we will describe the dynamical system for our interaction model.

\section{Dynamical system}

In this section, we will analyze the dynamic system of our model to determine its critical points,  along  with the cosmological parameters and stability of our autonomous system. In general terms,  
we introduce the following useful dimensionless variables,  defined as \cite{Copeland:2006wr}

\begin{eqnarray}
x =& \dfrac{\kappa  \dot{\phi}}{\sqrt{6} H}, \ \ \ \ \ \ y =& \dfrac{\kappa  \sqrt{V}}{\sqrt{3} H}, \ \ \ \ \ \ \ \varrho=\frac{\kappa\sqrt{\rho_r}}{\sqrt{3}H},\nonumber\\ 
u =& \dfrac{\kappa }{\sqrt{3 F} H}, \ \ \ \ \ \
\lambda =& - \dfrac{V_{,\phi}}{\kappa V}, \ \ \ \ \ \ \ \ \ \ \ \ \ \ \nonumber\\
\sigma =&  \dfrac{F_{,\phi}}{\kappa F},  \ \ \ \ \ \ \ \ \ \ \ \ \Gamma =&  \dfrac{V V_{,\phi\phi}}{(V_{,\phi})^2}, \ \ \ \ \ \ \ \Theta = \dfrac{F F_{,\phi\phi}}{(F_{,\phi})^2}. \nonumber\\
&& 
\label{var}
\end{eqnarray}
By using these variables the constraint equation defined by Eq.(\ref{ec}) can be rewritten as 
\begin{equation}
    \Omega_m + \varrho^2 +\frac{3 x^4}{u^2}+\alpha x^2+y^2= 1.
\end{equation}
Additionally, we can express the dynamical system in general terms using these dimensionless variables as follows
\begin{eqnarray}
\dfrac{d x}{d N} &=& f_1(x,y,u, \varrho),
\label{dinsyseq1}\\
\dfrac{d y}{d N} &=& f_2(x,y,u, \varrho),
\label{dinsyseq2}\\
\dfrac{d u}{d N} &=& f_3(x,y, u,\varrho),
\label{dinsyseq4}\\
\dfrac{d \varrho}{d N} &=& f_4(x,y,u, \varrho),
\label{dinsyseq5}\\
\dfrac{d \lambda}{d N} &=& -\sqrt{6} \left(\Gamma -1\right) \lambda ^2 x, \label{dinsyseq6}\\
\dfrac{d \sigma}{d N} &=& \sqrt{6} \left(\Theta -1\right) \sigma^2 x, \label{dinsyseq8}
\end{eqnarray}
where the functions $f_i(x,y,u, \varrho)$ with $i=1,2,3,4$  depend on the type of interaction to be studied. We  assume that the interaction term $Q$ can be expressed  in terms of these dimensionless variables. Besides, 
 the quantity $N$ corresponds to the number of $e-$folds, defined in terms of the scale factor as $N=\ln a$.

Additionally, considering the above set of phase space variables defined by Eq.(\ref{var}), we can express the density parameter associated  to the DE as
\bea
\Omega_{de} &=& 1 - \left(\frac{-\alpha  u^2 x^2 - u^2 y^2 + u^2 - 3 x^4}{u^2}\right)
. \nonumber\\
&&
\eea

Similarly, the parameter related to the equation of state related to the  DE is defined as $w_{de}=p_{de}/\rho_{de}$, which can be rewritten as
\bea
w_{de}&=& \frac{\alpha  u^2 x^2 - u^2 y^2 + x^4}{u^2 \left(1 - \frac{-\alpha  u^2 x^2 - u^2 y^2 + u^2 - 3 x^4}{u^2}\right)},
\eea
and in analogously to $w_{de}$, the total equation of state can be rewritten in terms of the new variables as
\bea
w_{tot} &=& \frac{x^4}{u^2} + \alpha x^2 - y^2 + \frac{\varrho^2}{3}.
\eea

In the following, we will analyze three different types of interactions arising from the source term $Q$,  and  we also present   
 a detailed dynamical analysis for these interaction models.

Examining the left-hand side of Eqs.(\ref{ecde}) and (\ref{rho_m}), it becomes clear that the interaction term
$Q$ must be a function of the energy densities $\rho_{de}$ or $\rho_{m}$  multiplied by a term with units of inverse time\cite{delCampo:2008jx,delCampo:2008sr,delCampo:2006vv,Herrera:2016uci}. In the literature, various combinations of these quantities are considered, such that the interaction term takes the form $Q=Q(\kappa\rho_m\dot{\phi},\kappa\rho_{de}\dot{\phi},H\rho_m,H\rho_{de},..)$. In this sense, we will study the three simplest interactions to analyze and the most commonly used in the literature, which are
\bea
Q=\beta\kappa\rho_m\dot{\phi},\,\,\,\,\,\,Q=3 \beta \rho_{m} H,\,\,\,\,\,\mbox{and}\,\,\,\,\,\,\,\,Q=3\beta\,\rho_{de}\, H,
\eea
in which $\beta$ is a constant and it corresponds to the dimensionless parameter. As mentioned above, different expressions for the interaction
$Q$ have been studied in the literature. For other forms of the interaction term $Q$, see  Refs.\cite{Amendola:1999er,Chimento:2003iea,Wang:2016lxa,He:2008tn,Yang:2018xlt,Herrera:2004dh,delCampo:2015vha,delCampo:2005tr}.

Additionally, to determine our autonomous system based on the dynamical system defined by Eqs.(\ref{dinsyseq1})-(\ref{dinsyseq8}), we need to specify  the effective potential $V(\phi)$ and the coupling function $F(\phi)$ in terms of the scalar field. In this context, and following Ref.\cite{Piazza:2004df}, we study a dark energy model with a ghost scalar field that interactions with the dark matter, in which the effective potential and the coupling function associated to the ghost scalar field are defined as
\bea
V(\phi)=V_0\,e^{-\lambda\,\phi}, \,\,\,\,\,\,\mbox{and}\,\,\,\,\,\,F(\phi)=F_0\,e^{\sigma\,\phi},\label{VF}
\eea
where $V_0$, $\lambda$, $F_0$ and $\sigma$ are constants. Here, the constant $V_0$ has dimension of $\kappa^{-4}=M_p^4$, the quantity $F_0$ of $\kappa^4=M_p^{-4}$ and the parameters $\lambda$ and $\sigma$ are positive quantities  and these parameters  have units of $\kappa=M_p^{-1}$.

\subsection{Interaction  $Q =\beta  \kappa \rho_m \dot\phi$. Critical points and Stability of critical points.}\label{phase_space}

In this section, we will describe the dynamical system 
for the first  interaction $Q\propto \rho_m\dot{\phi}$ together with the effective potential and the coupling function in terms of the scale factor defined by Eq.(\ref{VF}). In this context,  
 we find that the different functions $f_i(x,y,u,\varrho)$  describing  our dynamical system are given by 
\begin{eqnarray}
f_1 &=&\frac{u^4 \left[3 \alpha ^2 x^3 + \sqrt{6} \alpha  \beta  x^2 + \alpha  x \left(-3 y^2 + \varrho ^2 - 3\right) + \sqrt{6} \left(\beta  \left(y^2 + \varrho ^2 - 1\right) + \lambda  y^2 \right)\right]}{2 u^2 \left(\alpha  u^2 + 6 x^2\right)} \nonumber \\
&+& \frac{3 u^2 x^3 \left(7 \alpha  x^2 + \sqrt{6} x (\beta - \sigma) - 6 y^2 + 2 \left(\varrho ^2 + 1\right)\right) + 18 x^7}{2 u^2 \left(\alpha  u^2 + 6 x^2\right)},\label{f1} \\
f_2 &=&  \frac{1}{2} \left( \frac{3 x^4}{u^2} + 3 \alpha  x^2 - \sqrt{6} \lambda  x - 3 y^2 + \varrho ^2 + 3 \right)\,y,
 \\
f_3 &=&  \frac{3 x^4}{2 u} + \frac{1}{2} u \left(3 \alpha  x^2 - \sqrt{6} \sigma  x - 3 y^2 + \varrho ^2 + 3\right),
 \\
f_4 &=&  \frac{1}{2} \varrho  \left(\frac{3 x^4}{u^2} + 3 \alpha  x^2 - 3 y^2 + \varrho ^2 - 1\right).\label{f4}
 \\
\end{eqnarray}

\begin{table*}[ht]
 \centering
 \caption{Critical points for the autonomous system in the interaction $Q\propto\rho_m\,\dot{\phi} $. }
\begin{center}
\begin{tabular}{c c c c c c c c c}\hline\hline
Name &  $x_c$ & $y_c$ & $u_{c}$ & $\varrho_{c}$  \\\hline
$\ \ \ \ \ \ \ \ a_{R} \ \ \ \ \ \ \ \ $ & $0$ & $0$ & $0$  & $1$ \\
$\ \ \ \ \ \ \ \ b_{M} \ \ \ \ \ \ \ \ $ & $0$ & $0$ & $0$  & $0$ \\
$\ \ \ \ \ \ \ \ c \ \ \ \ \ \ \ \ $ & $ \frac{\sqrt{\frac{3}{2}}}{\beta +\sigma }$ & $0$ & $\frac{3}{\sqrt{2} \sqrt{-(\beta +\sigma )^2 (3 \alpha +2 \beta  (\beta +\sigma ))}}$  & $0$ \\
$\ \ \ \ \ \ \ \ d^{\pm} \ \ \ \ \ \ \ \ $ & $ \frac{16 \sqrt{6} \alpha  \sigma ^2 \pm \sqrt{2} \sqrt{-\sigma ^2 \left(16 \alpha -3 \sigma ^2\right)^3}-3 \sqrt{6} \sigma ^4}{64 \alpha ^2 \sigma -12 \alpha  \sigma ^3}$ & $0$ & $\frac{1}{2} \sqrt{\frac{96 \alpha ^2-72 \alpha  \sigma ^2 \mp \sqrt{3} \sqrt{\sigma ^2 \left(3 \sigma ^2-16 \alpha \right)^3}+9 \sigma ^4}{\alpha ^3 \left(6 \alpha -\sigma ^2\right)}}$  & $0$ \\

\\ \hline\hline
\end{tabular}
\end{center}
\label{table1}
\end{table*}
\begin{table}[ht]
 \centering
 \caption{Cosmological parameters for the critical points in Table \ref{table1}. }
\begin{center}
\begin{tabular}{c c c c c c}\hline\hline
Name &   $\Omega_{de}$ & $\Omega_{m}$ & $\Omega_{r}$ & $\omega_{de}$ & $\omega_{tot}$ \\\hline
$a_{R}$ & $0$ & $0$ & $1$ & $0$ & $\frac{1}{3}$ \\
$b_{M}$ & $0$ & $1$ & $0$ & $0$ & $0$ \\
$c$ & $-\frac{3 (\alpha +\beta  (\beta +\sigma ))}{(\beta +\sigma )^2}$ & $1+\frac{3 (\alpha +\beta  (\beta +\sigma ))}{(\beta +\sigma )^2}$ & $0$ & $\frac{\beta  (\beta +\sigma )}{3 (\alpha +\beta  (\beta +\sigma ))}$ & $-\frac{\beta }{\beta +\sigma}$ \\
$d^{\pm}$ & $1$ & $0$ & $0$ & $\frac{-96 \alpha ^2+66 \alpha  \sigma ^2 \pm \sqrt{3} \sqrt{\sigma ^2 \left(3 \sigma ^2-16 \alpha \right)^3}-9 \sigma ^4}{6 \alpha  \left(16 \alpha -3 \sigma ^2\right)}$ & $\frac{-96 \alpha ^2+66 \alpha  \sigma ^2 \pm \sqrt{3} \sqrt{\sigma ^2 \left(3 \sigma ^2-16 \alpha \right)^3}-9 \sigma ^4}{6 \alpha  \left(16 \alpha -3 \sigma ^2\right)}$ 
\\ \hline\hline
\end{tabular}
\end{center}
\label{table2}
\end{table}

Thus, from these functions $f_i$ we will find 
the critical points from our first interaction term, together with the potential and coupling function defined by Eq.(\ref{VF}). In this context,  the critical points    
are obtained by satisfying the conditions $dx/dN = dy/dN = du/dN=d\varrho/dN=0$ in Eqs.(\ref{dinsyseq1})-(\ref{dinsyseq8}), using the functions $f_i$ given by Eqs.(\ref{f1})-(\ref{f4}), respectively. Besides, considering the definition of the dynamical variables defined in Eq.(\ref{var}), the physically possible quantities associated with the critical points satisfy the conditions $y_c\ge 0$, $u_c\ge 0$ and $\varrho_c\ge 0$. Here, in the following the notation with the subscript   ``c'' denotes a critical point.  
Furthermore, the critical points of our system associated to the first source term $Q\propto\rho_m\dot{\phi}$ are shown in Table \ref{table1}. Furthermore, the different values of their cosmological parameters are presented in Table \ref{table2}. In these tables, the quantity $a_R$ corresponds to the critical point related to a radiation epoch. At this critical point, we have  $\Omega_r=1$, $w_{de}=0$ and the total EoS $w_{tot}=1/3$. This point is independent of the values of  $u_{c}$. Also, we note that these critical points do not depend on the parameter $\lambda$
 associated with effective potential.

The critical point $b_M$ corresponds to a matter-dominated era, where $\Omega_m=1$ and the parameters $w_{de}=w_{total}=0$. Moreover, the critical points $d^{\pm}$ represent dark energy-dominated solutions, both of which correspond to a de Sitter solution with EoS parameters $w_{de}=w_{tot}=-1$. In both cases, an accelerated expansion occurs for all parameter values.

The critical point $c$ in the special case where the parameter $\alpha= -\beta(\beta+\sigma)$ represents a matter-dominated era, in which $\Omega_m=1$ and $\Omega_{de}=0$. However, for this value, the EoS parameter associated with the dark energy, $w_{de}$, is not determined. In particular, for the case in which  $\alpha=-1$, we have the matter-dominated era takes place for values of $\beta$ given by $\beta=(1/2)[-\sigma\pm\sqrt{\sigma^2+4}]$.
In the case where $\beta=0$, along with the ratio $\alpha/\sigma^2\to 0$, we also obtain a matter-dominated era, in which $\Omega_m\to 1$ and $w_{de}=w_{tot}=0$. Furthermore, for a non-interacting model where $\beta=0$ and $\alpha/\sigma^2\neq 0$, we find a scaling matter epoch in which $\Omega_m=1+3\alpha/\sigma^2$ and $w_{de}=w_{tot}=0$.

On the other hand, to analyze the stability of the critical points, we consider small time-dependent linear perturbations in the dimensionless variables associated with the dynamical system around each critical point. In this sense, we can write $x=x_c+\delta x$, $y=y_c+\delta y$, $\varrho=\varrho_c+\delta\varrho$ and  $u=u_c+\delta u$. Here, the quantities $\delta x$, $\delta y$,  $\delta \varrho$ and  $\delta u$,  correspond to small perturbations, such that $x_c\gg\delta x$, $y_c\gg \delta y$, etc.

Thus, by introducing these perturbations into Eqs. (\ref{dinsyseq1})-(\ref{dinsyseq8}), we obtain the linear perturbation matrix denoted by $\mathcal{M}$ (see Ref. \cite{Copeland:2006wr}). In this way, we determine the eigenvalues of the matrix $\mathcal{M}$, which, when evaluated at each fixed point, are denoted as $\mu_i$ with $i=1,2,3,4$. Consequently, we can establish the stability conditions for the different points $\mu_i$.

In relation to the classification of the stability of critical points, we have a stable node when all the eigenvalues obtained are negative and an unstable node when all the eigenvalues found are positive. Additionally, in this classification, a saddle point occurs when the eigenvalues have different signs, and a stable spiral is obtained when the determinant of the matrix $\mathcal{M}$ is negative.
It is important to mention that the points classified as stable nodes or stable spirals correspond to attractor points \cite{Copeland:2006wr}.

In the following, we present the different eigenvalues and the stability conditions for the critical points obtained in relation to our interaction model.

\begin{itemize}
    \item Point $a_R$ has the eigenvalues   
    \begin{equation}
        \mu_1 = 2, \ \ \ \ \ \mu_2 = 2, \ \ \ \ \ \mu_3 = -1, \ \ \ \ \ \mu_4 = 1 ,
    \end{equation}
    therefore, it is a saddle point.
    \item Point $b_M$ has the eigenvalues
    \begin{equation}
        \mu_1 = -\dfrac{3}{2}, \ \ \ \ \ \mu_2 = \dfrac{3}{2}, \ \ \ \ \ \mu_3 = \dfrac{3}{2}, \ \ \ \ \ \mu_4 = -\dfrac{1}{2},
    \end{equation}
    therefore, it is a saddle point.
    \item Point $c$  has the eigenvalues
    \begin{eqnarray}
        \mu_1 &=& -\frac{4 \beta + \sigma}{2 (\beta + \sigma)}, \ \ \ \ \ \mu_2 = \frac{3 (\sigma - \lambda)}{2 (\beta + \sigma)},\nonumber\\
        \mu_3 &=& \frac{1}{4} \left( -\sqrt{3} \sqrt{-\frac{72 \alpha^2 + 3 \alpha (28 \beta^2 + 36 \beta \sigma + 3 \sigma^2) + 4 \beta (\beta + \sigma) (4 \beta^2 + 8 \beta \sigma + \sigma^2)}{(\beta + \sigma)^2 (5 \alpha + 4 \beta (\beta + \sigma))}} + \frac{3 \sigma}{\beta + \sigma} - 6 \right), \ \ \ \ \ \ \nonumber \\
        \mu_4 &=& \frac{1}{4} \left( \sqrt{3} \sqrt{-\frac{72 \alpha^2 + 3 \alpha (28 \beta^2 + 36 \beta \sigma + 3 \sigma^2) + 4 \beta (\beta + \sigma) (4 \beta^2 + 8 \beta \sigma + \sigma^2)}{(\beta + \sigma)^2 (5 \alpha + 4 \beta (\beta + \sigma))}} + \frac{3 \sigma}{\beta + \sigma} - 6 \right),
    \end{eqnarray}
    where, this critical point is a saddle point for
\begin{align}
  & (3 \alpha + 4 \beta^2 + 5 \beta \sigma + \sigma^2 \geq 0 \land (\alpha + \beta (\beta + \sigma) \leq 0)) \lor \nonumber \\
  & (\sigma \in \mathbb{R} \land ((\beta + \sigma > 0 \land \lambda < \sigma) \lor (\lambda > \sigma \land \beta + \sigma < 0))) \lor \nonumber \\
  & (\sigma < 0 \land \beta + \sigma < 0 \land (\sqrt{6} \sqrt{\sigma^2} + \sigma < 2 \beta \lor 4 \beta + \sigma > 0)) \lor \nonumber \\
  & (\sigma > 0 \land \beta + \sigma > 0 \land (4 \beta + \sigma < 0 \lor 2 \beta + \sqrt{6} \sqrt{\sigma^2} \leq \sigma)) ) \lor \nonumber \\
  & (\sigma < 0 \land (2 \beta + \sigma > 0 \land 8 \beta + 5 \sigma < 0 \land 5 \alpha + 4 \beta (\beta + \sigma) < 0) \nonumber \\
  & \lor (8 \beta + 5 \sigma > 0 \land \sqrt{(2 \beta + \sigma)^3 (34 \beta + 9 \sigma)} \geq 48 \alpha + 28 \beta^2 + 36 \beta \sigma + 3 \sigma^2 \land \nonumber \\
  & \sqrt{6} \sqrt{\sigma^2} + \sigma \geq 2 \beta)) ) \lor \nonumber \\
  & (\sigma > 0 \land ((8 \beta + 5 \sigma > 0 \land 2 \beta + \sigma < 0 \land 5 \alpha + 4 \beta (\beta + \sigma) < 0) \lor \nonumber \\
  & (2 \beta + \sqrt{6} \sqrt{\sigma^2} > \sigma \land \sqrt{(2 \beta + \sigma)^3 (34 \beta + 9 \sigma)} \geq 48 \alpha + 28 \beta^2 + 36 \beta \sigma + 3 \sigma^2 \land \nonumber \\
  & 8 \beta + 5 \sigma < 0))) ) \lor \nonumber \\
  & (5 \alpha + 4 \beta (\beta + \sigma) < 0 \land (\sigma > 0 \land ((2 \beta + \sigma = 0 \land 3 \alpha + 4 \beta^2 + 5 \beta \sigma + \sigma^2 > 0) \lor \nonumber \\
  & (2 \beta + \sigma > 0 \land 3 \alpha + 2 \beta (\beta + \sigma) > 0 \land \beta < 0))) \lor \nonumber \\
  & (\beta > 0 \land 3 \alpha + 2 \beta (\beta + \sigma) > 0 \land \sigma < 0 \land 2 \beta + \sigma \leq 0)) ).
\end{align}

    \item Points $d^{\pm}$ have the eigenvalues
\begin{eqnarray}
        \mu_1 &=& \frac{-128\alpha^2 + 72\alpha\sigma^2 - 9\sigma^4 \pm \sqrt{3}\sqrt{\sigma^2(-16\alpha + 3\sigma^2)^3}}{64\alpha^2 - 12\alpha\sigma^2},\nonumber\\
        \mu_2 &=& \frac{-192\alpha^2 + 84\alpha\sigma^2 - 9\sigma^4 \pm \sqrt{3}\sqrt{\sigma^2(-16\alpha + 3\sigma^2)^3}}{64\alpha^2 - 12\alpha\sigma^2}, \nonumber \\
        \mu_3 &=& \frac{(\lambda - \sigma) \left(-48 \alpha \sigma^2 + 9 \sigma^4 \pm \sqrt{3} \sqrt{\sigma^2 (-16 \alpha + 3 \sigma^2)^3}\right)}{4 \alpha \sigma (16 \alpha - 3 \sigma^2)},\nonumber \\
        \mu_4 &=& \frac{-96 \alpha^2 \sigma + 6 \alpha \sigma^2 (8 \beta + 11 \sigma) + (\beta + \sigma) \left(- 9 \sigma^4 \pm \sqrt{3} \sqrt{\sigma^2 (-16 \alpha + 3 \sigma^2)^3}\right)}{2 \alpha \sigma (16 \alpha - 3 \sigma^2)},
    \end{eqnarray}  
    
therefore, $d^-$ is stable for
   \begin{align}
    &\big((\alpha < 0 \lor \beta + \sigma > 0) \land (3\alpha + 4\beta^2 + 5\beta\sigma + \sigma^2 < 0 \lor \beta + \sigma \leq 0) \quad \land \sigma < 0 \land \lambda > \sigma\big) \nonumber \\
    &\lor \big((\alpha < 0 \lor \beta + \sigma < 0) \land (3\alpha + 4\beta^2 + 5\beta\sigma + \sigma^2 < 0 \lor \beta + \sigma \geq 0) \quad \land \lambda < \sigma \land \sigma > 0\big),
    \end{align}
    
   and $d^+$ is stable for
   \begin{align}
&\big((\alpha < 0 \lor 4\beta + \sigma \leq 0) \land 
(3\alpha + 4\beta^2 + 5\beta\sigma + \sigma^2 < 0 \lor 4\beta + \sigma > 0) \quad \land \lambda < \sigma \land \sigma < 0\big) \nonumber \\
&\lor \big((\alpha < 0 \lor 4\beta + \sigma > 0) \land 
(3\alpha + 4\beta^2 + 5\beta\sigma + \sigma^2 < 0 \lor 4\beta + \sigma \leq 0) \quad \land \lambda > \sigma \land \sigma > 0\big).
\end{align}

\end{itemize}

Subsequently, we validate our analytical results by numerically integrating the system of cosmological equations for our model. Additionally, we compare them with observational data.

 \subsection{Interaction  $Q =3 \beta \rho_m H$. Critical points and Stability of critical points }\label{phase_space}

In this section, we describe the dynamical system for the second interaction, in which the interaction term is given by $Q \propto \rho_m H$. For this interaction, we find that the different functions $f_i(x,y,u,\varrho)$, which describe our dynamical system given by Eqs. (\ref{dinsyseq1})–(\ref{dinsyseq8}), are given by

\begin{table*}[ht]
 \centering
 \caption{Critical points for the autonomous system in the interaction $Q\propto \rho_m\,H$. }
\begin{center}
\begin{tabular}{c c c c c c c c c}\hline\hline
Name &  $x_c$ & $y_c$ & $u_{c}$ & $\varrho_{c}$  \\\hline
$\ \ \ \ \ \ \ \ a_{R} \ \ \ \ \ \ \ \ $ & $0$ & $0$ & $0$  & $1$ \\
$\ \ \ \ \ \ \ \ b_{M} \ \ \ \ \ \ \ \ $ & $0$ & $0$ & $0$  & $0$ \\
$\ \ \ \ \ \ \ \ c \ \ \ \ \ \ \ \ $ & $ -\frac{\sqrt{\frac{3}{2}} (\beta -1)}{\sigma }$ & $0$ & $\frac{3 \sqrt{\frac{(\beta -1)^4 \left(\sigma ^2-6 \alpha \right)}{\sigma ^2}}}{\sqrt{2} \sqrt{-\left(\left(\sigma ^2-6 \alpha \right) \left(3 \alpha  (\beta -1)^2+2 \beta  \sigma ^2\right)\right)}}$  & $0$ \\
$\ \ \ \ \ \ \ \ d^{\pm} \ \ \ \ \ \ \ \ $ & $ \frac{\pm \sqrt{2} \sqrt{3 \sigma ^2-16 \alpha }+\sqrt{6} \sigma }{4 \alpha }$ & $0$ & $\frac{\sqrt[4]{3} \sqrt{\frac{\left(32 \sqrt{3} \alpha ^2-8 \alpha  \sigma  \left(\pm 2 \sqrt{3 \sigma ^2-16 \alpha }+3 \sqrt{3} \sigma \right)+3 \sigma ^3 \left(\pm \sqrt{3 \sigma ^2-16 \alpha }+\sqrt{3} \sigma \right)\right) \left(3 \alpha  (\beta -1)^2+2 \beta  \sigma ^2\right)}{\alpha ^2}}}{2 \sqrt{\alpha  \left(6 \alpha -\sigma ^2\right) \left(3 \alpha  (\beta -1)^2+2 \beta  \sigma ^2\right)}}$  & $0$ \\

\\ \hline\hline
\end{tabular}
\end{center}
\label{table3}
\end{table*}
\begin{table}[ht]
 \centering
 \caption{Cosmological parameters for the critical points in Table \ref{table3}. }
\begin{center}
\begin{tabular}{c c c c c c}\hline\hline
Name &   $\Omega_{de}$ & $\Omega_{m}$ & $\Omega_{r}$ & $\omega_{de}$ & $\omega_{tot}$ \\\hline
$a_{R}$ & $0$ & $0$ & $1$ & $0$ & $\frac{1}{3}$ \\
$b_{M}$ & $0$ & $1$ & $0$ & $0$ & $0$ \\
$c$ & $-\frac{3 (\alpha +\beta  (\beta +\sigma ))}{(\beta +\sigma )^2}$ & $1+\frac{3 (\alpha +\beta  (\beta +\sigma ))}{(\beta +\sigma )^2}$ & $0$ & $\frac{\beta  \sigma ^2}{3 \left(\alpha  (\beta -1)^2+\beta  \sigma ^2\right)}$ & $-\beta$ \\
$d^{\pm}$ & $-\frac{3 \alpha  (\beta -1)^2}{\sigma ^2} - 3 \beta $ & $\frac{3 \alpha  (\beta -1)^2}{\sigma ^2}+3 \beta +1$ & $0$ & $-1 \pm \frac{\sigma  \left(\sqrt{9 \sigma ^2-48 \alpha }+3 \sigma \right)}{6 \alpha }$ & $-1 \pm \frac{\sigma  \left(\sqrt{9 \sigma ^2-48 \alpha }+3 \sigma \right)}{6 \alpha }$ 
\\ \hline\hline
\end{tabular}
\end{center}
\label{table4}
\end{table}

\begin{eqnarray}
f_1 &=&\frac{u^4 \left(3 \alpha ^2 x^4 + \alpha  x^2 \left(3 \beta - 3 y^2 + \varrho^2 - 3\right) + \sqrt{6} \lambda  x y^2 + 3 \beta  \left(y^2 + \varrho^2 - 1\right)\right)}{2 u^2 \left(\alpha  u^2 x + 6 x^3\right)} \nonumber \\
&+& \frac{3 u^2 x^4 \left(3 \beta + 7 \alpha  x^2 - \sqrt{6} \sigma x - 6 y^2 + 2 \varrho^2 + 2\right) + 18 x^8}{2 u^2 \left(\alpha  u^2 x + 6 x^3\right)},\label{fq1} \\
f_2 &=&  \frac{1}{2} \left( \frac{3 x^4}{u^2} + 3 \alpha  x^2 - \sqrt{6} \lambda  x - 3 y^2 + \varrho ^2 + 3 \right)\,y,
 \\
f_3 &=&  \frac{3 x^4}{2 u} + \frac{1}{2} u \left(3 \alpha  x^2 - \sqrt{6} \sigma  x - 3 y^2 + \varrho ^2 + 3\right),
 \\
f_4 &=&  \frac{1}{2} \varrho  \left(\frac{3 x^4}{u^2} + 3 \alpha  x^2 - 3 y^2 + \varrho ^2 - 1\right).\label{fq2}
\end{eqnarray}

Thus, as before, we determine the critical points for our second interaction term $Q$, along with the potential and coupling function defined by Eq. (\ref{VF}). As we have seen previously, the critical points are found by satisfying the conditions $dx/dN = dy/dN = du/dN = d\varrho/dN = 0$ in Eqs. (\ref{dinsyseq1})–(\ref{dinsyseq8}), using the functions provided in Eqs. (\ref{fq1})–(\ref{fq2}).

As before, we now present the eigenvalues and stability conditions for the critical points associated with our second interaction model. In this context, the quantities physically admissible at the critical points must satisfy conditions $y_c \geq 0$, $u_c \geq 0$, and $\varrho_c \geq 0$. The critical points of our system for the second source term, $Q \propto \rho_m H$, are listed in Table \ref{table3}, while the corresponding values of their cosmological parameters are shown in Table \ref{table4}.

 As before, the critical point $a_R$ describes a radiation-dominated epoch, where the parameter $\Omega_r = 1$ and the total equation of state (EoS) parameter is $w_{tot} = 1/3$. The point $b_M$ represents a matter-dominated epoch, characterized by $\Omega_m = 1$ and EoS parameters $w_{de} = 0$ and $w_{tot} = 0$, respectively.

The critical point $c$ corresponds to a scaling matter era, where the dark energy density parameter is given by $\Omega_{de} = -3[\alpha + \beta(\beta + \sigma)]/(\beta + \sigma)^2$. In the absence of interaction ($\beta = 0$) and under the condition $\sigma \to \pm \infty$, the standard matter-dominated scenario is recovered. In particular, the critical point $c$ arises from both the interaction term and the coupling function of the dilaton field associated with the higher-order kinetic term.

Furthermore, the total EoS parameter depends only on the interaction term through the parameter $\beta$. Specifically, for $\beta > 1/3$, the universe undergoes an accelerated expansion.

Additionally, the critical points $d^{\pm}$ correspond to a matter-dominated era driven by both the interaction term and the coupling function of the dilaton field. In particular, when the interaction parameter satisfies $\beta = [(2 - \sigma^2/\alpha) \pm \sqrt{\sigma^2/\alpha - 4}, \sigma \alpha^{-1/2}]/2$, the standard matter-dominated era is recovered.

Finally, we observe that the EoS parameters are independent of the interaction term $\beta$. In the absence of a coupling function for the dilaton field associated with the higher-order kinetic term, the system admits a dark energy-dominated solution with an EoS of $w_{de} = w_{tot} = -1$, corresponding to a de Sitter accelerated solution.

On the other hand, to analyze the stability of the critical points, we follow the same approach as before by considering small time-dependent linear perturbations in the dimensionless variables of the dynamical system around each critical point for the second interaction term. In this context, we now present the eigenvalues and stability conditions for the critical points obtained from the second interaction term, given by $Q \propto \rho_m H$.

\begin{itemize}
    \item Point $a_R$ has the eigenvalues   
    \begin{equation}
        \mu_1 = 2, \ \ \ \ \ \mu_2 = 2, \ \ \ \ \ \mu_3 = -1, \ \ \ \ \ \mu_4 = 1, 
    \end{equation}
    therefore, it is a saddle point.
    \item Point $b_M$ has the eigenvalues
    \begin{equation}
        \mu_1 = -\dfrac{3}{2}, \ \ \ \ \ \mu_2 = \dfrac{3}{2}, \ \ \ \ \ \mu_3 = \dfrac{3}{2}, \ \ \ \ \ \mu_4 = -\dfrac{1}{2},
    \end{equation}
    therefore, it is a saddle point.
    \item Point $c$  has the eigenvalues
    \begin{eqnarray}
        \mu_1 &=& \frac{1}{2} (-1 - 3 \beta), \ \ \ \ \ \mu_2 = \frac{3 (\beta - 1) (\lambda - \sigma)}{2 \sigma},\nonumber\\
        \mu_3 &=& -\frac{15 \alpha - 9 \alpha \beta - 27 \alpha \beta^2 + 21 \alpha \beta^3 + 14 \beta \sigma^2 + 
18 \beta^2 \sigma^2 + 5 \alpha \sqrt{A_1} - 10 \alpha \beta \sqrt{A_1} + 
5 \alpha \beta^2 \sqrt{A_1} + 4 \beta \sigma^2 \sqrt{A_1}}{4 \left( 5 \alpha (\beta - 1)^2 + 4 \beta \sigma^2 \right)}, \ \ \ \ \ \ \nonumber \\
        \mu_4 &=& \frac{-15 \alpha + 9 \alpha \beta + 27 \alpha \beta^2 - 21 \alpha \beta^3 - 14 \beta \sigma^2 - 
18 \beta^2 \sigma^2 + 5 \alpha \sqrt{A_1} - 10 \alpha \beta \sqrt{A_1} + 
5 \alpha \beta^2 \sqrt{A_1} + 4 \beta \sigma^2 \sqrt{A_1}}{4 \left( 5 \alpha (\beta - 1)^2 + 4 \beta \sigma^2 \right)}
    \end{eqnarray}
    where the quantity $A_1$ is defined as 
   \begin{align}
A_1 &= \frac{
1080 \alpha^3 (\beta - 1)^7 
+ 27 \alpha^2 (\beta - 1)^4 (-5 - 62 \beta + 115 \beta^2) \sigma^2 
}{\sigma^2 \left( 5 \alpha (\beta - 1)^2 + 4 \beta \sigma^2 \right)^2} \nonumber \\
&\quad + \frac{
36 \alpha (\beta - 1)^2 \beta (-3 - 14 \beta + 81 \beta^2) \sigma^4
+ 4 \beta^2 (1 + 15 \beta)^2 \sigma^6
}{\sigma^2 \left( 5 \alpha (\beta - 1)^2 + 4 \beta \sigma^2 \right)^2}.
\end{align}
    This critical point is a saddle point for
    \begin{align}
 & \sigma > 0 \; \wedge \Bigg(
    \Big(
        \beta < -\frac{1}{3} \wedge 
        -\frac{(1 + 3 \beta) \sigma^2}{3 (-1 + \beta)^2} \leq \alpha \leq 
        -\frac{\beta \sigma^2}{(-1 + \beta)^2} \wedge \lambda > 0 \wedge A_1 > 0
    \Big) \nonumber \\
    & \vee 
    \Big(
        \beta = -\frac{1}{3} \wedge 
        \Big(
            0 \leq \alpha < \frac{3 \sigma^2}{20} \wedge 
            \Big(
                0 < \lambda < \sigma \wedge A_1 > 0 
                \vee 
                \lambda \geq \sigma \wedge 
                A_1 > \frac{4 (16 \alpha - 3 \sigma^2)^2}{(20 \alpha - 3 \sigma^2)^2}
            \Big)
        \Big) \nonumber \\
    & \quad \vee 
        \Big(
            20 \alpha = 3 \sigma^2 \wedge 
            0 < \lambda < \sigma \wedge A_1 > 0
        \Big)  \quad \vee 
        \Big(
            \frac{3 \sigma^2}{20} < \alpha \leq \frac{3 \sigma^2}{16} \wedge 
            \lambda > 0 \wedge A_1 > 0
        \Big)
    \Big)
    \nonumber \\
    & \vee 
    \Big(
        -\frac{1}{3} < \beta \leq 0 \wedge 
        \Big(
            A_1 > 0 \wedge 
            \Big(
                0 < \lambda < \sigma \wedge 
                \Big(
                    \alpha = -\frac{4 \beta \sigma^2}{5 (-1 + \beta)^2} 
                    \vee 
                    -\frac{(1 + 3 \beta) \sigma^2}{3 (-1 + \beta)^2} \leq \alpha < 
                    -\frac{4 \beta \sigma^2}{5 (-1 + \beta)^2}
                    \nonumber \\
                    & \quad \vee 
                    -\frac{2 \beta (7 + 9 \beta) \sigma^2}{3 (-1 + \beta)^2 (5 + 7 \beta)} 
                    < \alpha \leq -\frac{\beta \sigma^2}{(-1 + \beta)^2}
                \Big)
            \Big)
            \vee 
            \Big(
                \lambda > 0 \wedge 
                -\frac{4 \beta \sigma^2}{5 (-1 + \beta)^2} < \alpha \leq 
                -\frac{2 \beta (7 + 9 \beta) \sigma^2}{3 (-1 + \beta)^2 (5 + 7 \beta)}
            \Big)
        \Big) \nonumber \\
        & \quad \vee 
        \Big(
            A_1 > \frac{(3 \alpha (-1 + \beta)^2 (5 + 7 \beta) + 
            2 \beta (7 + 9 \beta) \sigma^2)^2}{(5 \alpha (-1 + \beta)^2 + 4 \beta \sigma^2)^2} 
            \wedge 
            \lambda \geq \sigma \wedge 
            \Big(
                -\frac{(1 + 3 \beta) \sigma^2}{3 (-1 + \beta)^2} \leq \alpha < 
                -\frac{4 \beta \sigma^2}{5 (-1 + \beta)^2} 
                \nonumber \\
                & \quad \vee 
                -\frac{2 \beta (7 + 9 \beta) \sigma^2}{3 (-1 + \beta)^2 (5 + 7 \beta)} 
                < \alpha \leq -\frac{\beta \sigma^2}{(-1 + \beta)^2}
            \Big)
        \Big)
    \Big) \nonumber \\
    & \vee 
    \Big(
        0 < \beta < 1 \wedge 
        -\frac{(1 + 3 \beta) \sigma^2}{3 (-1 + \beta)^2} \leq \alpha \leq 
        -\frac{\beta \sigma^2}{(-1 + \beta)^2} \wedge 
        \Big(
            0 < \lambda < \sigma \wedge A_1 > 0 
            \nonumber \\
            & \quad \vee 
            \lambda \geq \sigma \wedge 
            A_1 > \frac{(3 \alpha (-1 + \beta)^2 (5 + 7 \beta) + 
            2 \beta (7 + 9 \beta) \sigma^2)^2}{(5 \alpha (-1 + \beta)^2 + 4 \beta \sigma^2)^2}
        \Big)
    \Big) \nonumber \\
    & \vee 
    \Big(
        \beta > 1 \wedge 
        -\frac{(1 + 3 \beta) \sigma^2}{3 (-1 + \beta)^2} \leq \alpha \leq 
        -\frac{\beta \sigma^2}{(-1 + \beta)^2} \wedge 
        \Big(
            0 < \lambda \leq \sigma \wedge 
            A_1 > \frac{(3 \alpha (-1 + \beta)^2 (5 + 7 \beta) + 
            2 \beta (7 + 9 \beta) \sigma^2)^2}{(5 \alpha (-1 + \beta)^2 + 4 \beta \sigma^2)^2} 
            \nonumber \\
            & \quad \vee 
            \lambda > \sigma \wedge A_1 > 0
        \Big)
    \Big)
\Bigg).
\end{align}
    \item Points $d^{\pm}$ have the eigenvalues
     \begin{align}
\mu_1 &= \frac{-192\alpha^2 + 84\alpha\sigma^2 - 9\sigma^4 \pm 
\sqrt{3} \sqrt{\sigma^2 (-16\alpha + 3\sigma^2)^3}}{4\alpha (16\alpha - 3\sigma^2)}, \nonumber \\
\mu_2 &= \frac{-128\alpha^2 + 72\alpha\sigma^2 - 9\sigma^4 \pm 
\sqrt{3} \sqrt{\sigma^2 (-16\alpha + 3\sigma^2)^3}}{4\alpha (16\alpha - 3\sigma^2)}, \nonumber \\
\mu_3 &= \frac{(\lambda - \sigma)(-48\alpha\sigma^2 + 9\sigma^4 \mp 
\sqrt{3} \sqrt{\sigma^2 (-16\alpha + 3\sigma^2)^3})}{4\alpha \sigma (16\alpha - 3\sigma^2)}, \nonumber \\
\mu_4 &= \frac{-96\alpha^2\sigma + 6\alpha\sigma^2(8\beta + 11\sigma) + 
(\beta + \sigma)(-9\sigma^4 \pm \sqrt{3} \sqrt{\sigma^2 (-16\alpha + 3\sigma^2)^3})}{2\alpha\sigma (16\alpha - 3\sigma^2)},
\end{align}
    therefore, $d^-$ is stable for
   \begin{align}
    &\big((\alpha < 0 \lor \beta + \sigma > 0) \land (3\alpha + 4\beta^2 + 5\beta\sigma + \sigma^2 < 0 \lor \beta + \sigma \leq 0) \quad \land \sigma < 0 \land \lambda > \sigma\big) \nonumber \\
    &\lor \big((\alpha < 0 \lor \beta + \sigma < 0) \land (3\alpha + 4\beta^2 + 5\beta\sigma + \sigma^2 < 0 \lor \beta + \sigma \geq 0) \quad \land \lambda < \sigma \land \sigma > 0\big).
    \end{align}
   Besides, the point  $d^+$ is stable for
   \begin{align}
&\big((\alpha < 0 \lor 4\beta + \sigma \leq 0) \land 
(3\alpha + 4\beta^2 + 5\beta\sigma + \sigma^2 < 0 \lor 4\beta + \sigma > 0) \quad \land \lambda < \sigma \land \sigma < 0\big) \nonumber \\
&\lor \big((\alpha < 0 \lor 4\beta + \sigma > 0) \land 
(3\alpha + 4\beta^2 + 5\beta\sigma + \sigma^2 < 0 \lor 4\beta + \sigma \leq 0) \quad \land \lambda > \sigma \land \sigma > 0\big).
\end{align}

\end{itemize}

\subsection{Interaction  $Q =3 \beta \rho_{de} H$. Critical points and stability of critical points.}\label{phase_space}

In this subsection, we describe the dynamical system for the third interaction, defined as $Q \propto \rho_{de} H$. For this interaction term $Q$, we find that the functions $f_i(x,y,u,\varrho)$ that characterize our dynamical system are given by

\begin{eqnarray}
f_1 &=&\frac{u^4 \left(3 \alpha ^2 x^4 - \alpha x^2 \left(\beta + 3 y^2 - \varrho^2 + 3\right) + \sqrt{6} \lambda x y^2 - \beta y^2 \right)}{2 u^2 \left(\alpha u^2 x + 6 x^3\right)} \nonumber \\
&+& \frac{3 u^2 x^4 \left(-\beta + 7 \alpha x^2 - \sqrt{6} \sigma x - 6 y^2 + 2 \varrho^2 + 2\right) + 18 x^8}{2 u^2 \left(\alpha u^2 x + 6 x^3\right)} \label{fq3},\\
f_2 &=&  \frac{1}{2} \left( \frac{3 x^4}{u^2} + 3 \alpha  x^2 - \sqrt{6} \lambda  x - 3 y^2 + \varrho ^2 + 3 \right)\,y,
 \\
f_3 &=&  \frac{3 x^4}{2 u} + \frac{1}{2} u \left(3 \alpha  x^2 - \sqrt{6} \sigma  x - 3 y^2 + \varrho ^2 + 3\right),
 \\
f_4 &=&  \frac{1}{2} \varrho  \left(\frac{3 x^4}{u^2} + 3 \alpha  x^2 - 3 y^2 + \varrho ^2 - 1\right)\label{fq4}.
 \\
\end{eqnarray}


As before, we determine the critical points for the interaction term $Q \propto \rho_{de} H$, along with the potential and coupling function defined by Eq. (\ref{VF}). The critical points are obtained by solving the conditions $dx/dN = dy/dN = du/dN = d\varrho/dN = 0$, using the functions $f_i$ given in Eqs. (\ref{fq3})–(\ref{fq4}).

Thus, we now present the eigenvalues and stability conditions for the critical points corresponding to the interaction term $Q \propto \rho_{de} H$. Physically admissible quantities associated with critical points must satisfy conditions $y_c \geq 0$, $u_c \geq 0$, and $\varrho_c \geq 0$. The critical points of our system, corresponding to the source term $Q \propto \rho_{de} H$, are summarized in Table \ref{table5}. Moreover, the corresponding values of their cosmological parameters are provided in Table \ref{table6}.

As mentioned earlier, the critical point $a_R$ describes a radiation-dominated epoch where the parameter $\Omega_r = 1$ and the total equation of state  parameter $w_{tot} = 1/3$. The point $b_M$ corresponds to a matter-dominated epoch where $\Omega_m = 1$, and the respective parameters $w_{de}$ and $w_{tot}$ are both zero. The critical point $c$ corresponds to a radiation-dominated scenario, where $\Omega_r = 1$ and $w_{tot} = 1/3$. In particular, the dark energy parameter $w_{de}$ depends on the value of $\beta$, which is related to the interaction term $Q$.
\\
\\

\begin{table*}[ht]
 \centering
 \caption{Critical points for the autonomous system in the interaction $Q\propto\rho_{de}\,H$.  Definitions of $x_{0}^{\pm}$ and $u_{0}^{\pm}$ are given at Appendix  \ref{AppB}}
\begin{center}
\begin{tabular}{c c c c c c c c c}\hline\hline
Name &  $x_c$ & $y_c$ & $u_{c}$ & $\varrho_{c}$  \\\hline
$\ \ \ \ \ \ \ \ a_{R} \ \ \ \ \ \ \ \ $ & $0$ & $0$ & $0$  & $1$ \\
$\ \ \ \ \ \ \ \ b_{M} \ \ \ \ \ \ \ \ $ & $0$ & $0$ & $0$  & $0$ \\
$\ \ \ \ \ \ \ \ c \ \ \ \ \ \ \ \ $ & $ \frac{2 \sqrt{\frac{2}{3}}}{\sigma }$ & $0$ & $\frac{2 \sqrt{2} \sqrt{\beta }}{\sqrt{-\alpha  (\beta +2) \sigma ^2}}$  & $\frac{\sqrt{3 \beta  \sigma ^2-16 \alpha  (\beta -1)}}{\sqrt{3} \sqrt{\beta } \sigma }$ \\
$\ \ \ \ \ \ \ \ d^{\pm} \ \ \ \ \ \ \ \ $ & $ x_{0}^{\pm}$ & $0$ & $u_{0}^{\pm}$  & $0$ \\

\\ \hline\hline
\end{tabular}
\end{center}
\label{table5}
\end{table*}
\begin{table}[ht]
 \centering
 \caption{Cosmological parameters for the critical points in Table \ref{table5}. }
\begin{center}
\begin{tabular}{c c c c c c}\hline\hline
Name &   $\Omega_{de}$ & $\Omega_{m}$ & $\Omega_{r}$ & $\omega_{de}$ & $\omega_{tot}$ \\\hline
$a_{R}$ & $0$ & $0$ & $1$ & $0$ & $\frac{1}{3}$ \\
$b_{M}$ & $0$ & $1$ & $0$ & $0$ & $0$ \\
$c$ & $-\frac{16 \alpha }{3 \beta  \sigma ^2}$ & $\frac{16 \alpha }{3 \sigma ^2}$ & $1-\frac{16 \alpha  (\beta -1)}{3 \beta  \sigma ^2}$ & $\frac{1-\beta }{3}$ & $\frac{1}{3}$ \\
$d^{\pm}$ & $-\frac{3 \alpha  (\beta -1)^2}{\sigma ^2} - 3 \beta $ & $\frac{3 \alpha  (\beta -1)^2}{\sigma ^2}+3 \beta +1$ & $0$ & $-1 \pm \frac{\sigma  \left(\sqrt{9 \sigma ^2-48 \alpha }+3 \sigma \right)}{6 \alpha }$ & $-1 \pm \frac{\sigma  \left(\sqrt{9 \sigma ^2-48 \alpha }+3 \sigma \right)}{6 \alpha }$ 
\\ \hline\hline
\end{tabular}
\end{center}
\label{table6}
\end{table}


On the other hand, as mentioned earlier, to analyze the stability of the critical points, we consider small time-dependent linear perturbations in the dimensionless variables of the dynamical system around each critical point. In this context, we will present the eigenvalues and stability conditions for the critical points obtained, considering the third interaction term given by $Q \propto \rho_{de} H$.

\begin{itemize}
    \item Point $a_R$ has the eigenvalues   
    \begin{equation}
        \mu_1 = 2, \ \ \ \ \ \mu_2 = 2, \ \ \ \ \ \mu_3 = -1, \ \ \ \ \ \mu_4 = 1, 
    \end{equation}
    therefore, it is a saddle point.
    \item Point $b_M$ has the eigenvalues
    \begin{equation}
        \mu_1 = -\dfrac{3}{2}, \ \ \ \ \ \mu_2 = \dfrac{3}{2}, \ \ \ \ \ \mu_3 = \dfrac{3}{2}, \ \ \ \ \ \mu_4 = -\dfrac{1}{2}.
    \end{equation}
    \item Points $d^{\pm}$ have the eigenvalues
     \begin{align}
\mu_1 &= \frac{-192\alpha^2 + 84\alpha\sigma^2 - 9\sigma^4 \pm 
\sqrt{3} \sqrt{\sigma^2 (-16\alpha + 3\sigma^2)^3}}{4\alpha (16\alpha - 3\sigma^2)}, \nonumber \\
\mu_2 &= \frac{-128\alpha^2 + 72\alpha\sigma^2 - 9\sigma^4 \pm 
\sqrt{3} \sqrt{\sigma^2 (-16\alpha + 3\sigma^2)^3}}{4\alpha (16\alpha - 3\sigma^2)}, \nonumber \\
\mu_3 &= \frac{(\lambda - \sigma)(-48\alpha\sigma^2 + 9\sigma^4 \mp 
\sqrt{3} \sqrt{\sigma^2 (-16\alpha + 3\sigma^2)^3})}{4\alpha \sigma (16\alpha - 3\sigma^2)}, \nonumber \\
\mu_4 &= \frac{-96\alpha^2\sigma + 6\alpha\sigma^2(8\beta + 11\sigma) + 
(\beta + \sigma)(-9\sigma^4 \pm \sqrt{3} \sqrt{\sigma^2 (-16\alpha + 3\sigma^2)^3})}{2\alpha\sigma (16\alpha - 3\sigma^2)},
\end{align}
    therefore, $d^-$ is stable for
   \begin{align}
    &\big((\alpha < 0 \lor \beta + \sigma > 0) \land (3\alpha + 4\beta^2 + 5\beta\sigma + \sigma^2 < 0 \lor \beta + \sigma \leq 0) \quad \land \sigma < 0 \land \lambda > \sigma\big) \nonumber \\
    &\lor \big((\alpha < 0 \lor \beta + \sigma < 0) \land (3\alpha + 4\beta^2 + 5\beta\sigma + \sigma^2 < 0 \lor \beta + \sigma \geq 0) \quad \land \lambda < \sigma \land \sigma > 0\big),
    \end{align}
    and $d^+$ is stable for
   \begin{align}
&\big((\alpha < 0 \lor 4\beta + \sigma \leq 0) \land 
(3\alpha + 4\beta^2 + 5\beta\sigma + \sigma^2 < 0 \lor 4\beta + \sigma > 0) \quad \land \lambda < \sigma \land \sigma < 0\big) \nonumber \\
&\lor \big((\alpha < 0 \lor 4\beta + \sigma > 0) \land 
(3\alpha + 4\beta^2 + 5\beta\sigma + \sigma^2 < 0 \lor 4\beta + \sigma \leq 0) \quad \land \lambda > \sigma \land \sigma > 0\big),
\end{align}
and  therefore it is a saddle point.
    
\end{itemize}

\section{Numerical Results}\label{Num_Res}

In this section, we aim to numerically solve the autonomous system defined by Eqs. \eqref{dinsyseq1}–\eqref{dinsyseq5}. We analyze the characteristics of our different interacting models to explain the current accelerated expansion of the universe. Furthermore, we compared the results obtained with the most recent observational data from $H(z)$ and Type Ia supernovae (SNe Ia). We will begin by analyzing the Hubble parameter as a function of the redshift $z$, i.e., $H(z)$, and then proceed to analyze the supernova Ia data.

\subsection{Hubble parameter H(z)}

To analyze the behavior of the Hubble parameter as a function of redshift and its confidence interval for the different interaction models, we will use a set of 39 data points provided by \cite{Farooq:2016zwm,Ryan:2018aif}, as detailed in Table \ref{table:H(z)data} (see Appendix \ref{appen_B}).

Additionally, for comparison purposes, we consider the $\Lambda$CDM model, which gives us the Hubble rate as a function of redshift:
\be
H_{\Lambda CDM}(z)=H_{0}\sqrt{\Omega_{de}^{(0)}+\Omega_{m}^{(0)}(1+z)^3+\Omega_{r}^{(0)}(1+z)^4} ,\label{HE}
\ee
where the subscript $(0)$ denotes the present value of the respective  density parameter.

\subsection{Supernovae Ia}

To incorporate the information from Supernova Ia, we examine the luminosity distance $D_L(z)$ as a function of redshift in a flat FLRW universe, as defined by \cite{Copeland:2006wr}:
\be
D_{L}(z)=\frac{1+z}{H_{0}}\int_{0}^{z} \frac{1}{h(z')} dz',
\ee 
where the dimensionless quantity $h(z)$ is defined as $h(z) \equiv H(z)/H_0$. Note that the above equation can also be expressed in differential form as:
\be
\frac{dD_{L}(z)}{dz}-\frac{D_{L}(z)}{1+z}-\frac{1+z}{H(z)}=0. \label{dD}
\ee 
Equation (\ref{dD}) is useful to integrate $D_L(z)$ when an analytical solution for $H(z)$ is not available.

In relation to the definition of the luminosity distance, we can analyze the difference between the apparent magnitude $m$ of the source and its absolute magnitude $M$, a quantity known as the distance modulus: 
\be \mu(z) \equiv m - M = 5 \log_{10}\left(\frac{D_L(z)}{M_{pc}}\right) + 25. \label{mu} 
\ee
Here, the numerical factor arises from the conventional definitions of $m$ and $M$ in astronomy \cite{Copeland:2006wr,amendola2010dark}.

In the following, we will numerically analyze the solutions to the autonomous system, along with the corresponding observational parameters, for the different interactions studied.

\subsubsection{Interaction $Q=\beta\kappa\rho_m\dot{\phi}$}

\begin{figure}[!tbp]
  \centering
    \includegraphics[width=0.5\linewidth]{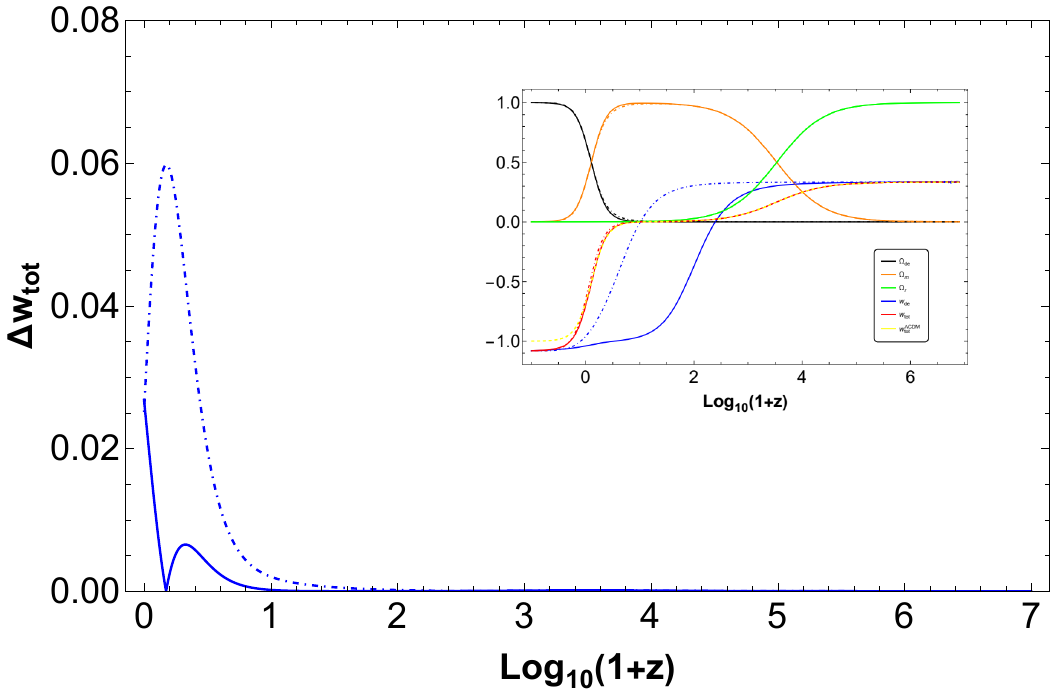} 
    \includegraphics[width=0.5\linewidth]{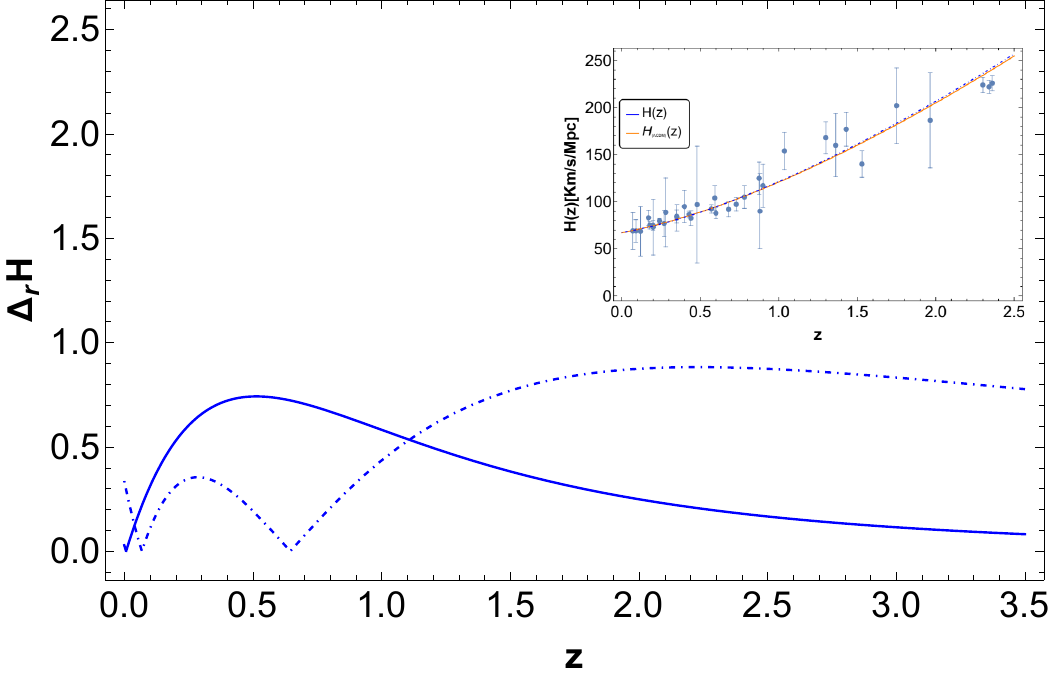} 
    \caption{\scriptsize{The upper panel displays the evolution of the relative difference $\Delta w_{tot}$ associated with the total EoS parameter as a function of Log${10}(1+z)$. Additionally, this figure shows the evolution of the various density parameters, along with the EoS parameters $w_{de}$, $w_{tot}$, and $w_{tot}^{\Lambda CDM}$,  as functions of Log$_{10}(1+z)$.
The lower panel illustrates the development of the relative difference $\Delta_r H$ related to the Hubble parameter as a function of the redshift $z$. Moreover, the inset in this panel displays the Hubble parameter $H(z)$ obtained from our interaction model, alongside the Hubble parameter for the $\Lambda$CDM model ($H_{(\Lambda CDM)}(z)$, as given by Eq. (\ref{HE})), compared against the observational data as a function of redshift.
In both panels, two different sets of initial variables $x_i$, $y_i$, $u_i$, and $\varrho_i$ are considered for comparison.
    }
    } 
    \label{Fig1}
\end{figure}

In this subsection, we present the numerical results for our first interaction, where the interaction term is given by $Q \propto \rho_m \dot{\phi}$. In this context, the upper panel of Fig. \ref{Fig1} shows the evolution of the absolute difference $\Delta w_{tot}$, associated with the total equation of state (EoS) parameter, defined as $\Delta w_{tot} = \left| w_{tot} - w_{tot}^{\Lambda \text{CDM}} \right|$, in terms of the function $\log_{10}(1+z)$.

We consider two different sets of values for the parameters and initial conditions, related to the values of $\lambda$, $\sigma$, $\alpha$, $\beta$, and the initial variables $x_i$, $y_i$, $u_i$, and $\varrho_i$, respectively. The dot-dashed line corresponds to the parameter values $\lambda = 0.1$, $\sigma = 0.1$, $\alpha = -1$, $\beta = 1.0 \times 10^{-1}$, and the initial variables $x_i = 1.0 \times 10^{-11}$, $y_i = 5.0 \times 10^{-13}$, $u_i = 9.5 \times 10^{-13}$, and $\varrho_i = 0.99983$. The solid line is associated with the parameter values $\lambda = 10^{-3}$, $\sigma = 0.1$, $\alpha = -1$, $\beta = 1.0 \times 10^{-3}$, and the initial variables $x_i = 1.0 \times 10^{-11}$, $y_i = 5.0 \times 10^{-13}$, $u_i = 1.25 \times 10^{-12}$, and $\varrho_i = 0.99983$.

From this graph, we observe that at the present time, where $z = 0$, the related difference is $\Delta w_{tot}(z=0) \ll 1$, suggesting that $w_{tot}$, obtained by considering the first interaction, is similar to $w_{tot}^{\Lambda \text{CDM}}$. That is, $w_{tot} \simeq w_{tot}^{\Lambda \text{CDM}}$.
Moreover, we observe that the maximum value of the absolute difference occurs at an approximate value of $\log_{10}(1+z) \simeq 0.5$ for the initial variables corresponding to the dot-dashed line, and at $z = 0$ for the initial variables associated with the solid line. Additionally, in this figure, we show the evolution of the different density parameters along with the EoS parameters $w_{de}$, $w_{tot}$, and $w_{tot}^{\Lambda \text{CDM}}$ as functions of $\log_{10}(1+z)$.

Specifically, within the upper panel, we depict the evolution of the fractional energy densities of dark energy $\Omega_{de}$ (black), dark matter (including baryons) $\Omega_m$ (orange), radiation $\Omega_r$ (green), the equation of state parameter of dark energy $w_{de}$ (blue), the total EoS parameter $w_{tot}$ (red), and the EoS parameter of the $\Lambda$CDM model (yellow) as functions of the cosmological redshift.

From this panel, we note that the EoS parameter $w_{tot}$ becomes negative for values of $\log_{10}(1+z) \lesssim 1$, and the accelerated expansion occurs when this parameter drops below $-1/3$. Specifically, we find that at $z = 0$, the EoS parameter associated with dark energy, $w_{de}$, takes the value $w_{de} = -1.039$, regardless of the initial conditions (dot-dashed line) related to the autonomous system. Furthermore, we find that the total EoS parameter at $z = 0$ is $w_{tot} \simeq -0.707$, a value that is close to that of the $\Lambda$CDM model, as seen from the figure of $\Delta w_{tot}$, where $\Delta w_{tot} \simeq 0.027$.

In addition, the lower panel of Fig. \ref{Fig1} shows the relative difference $\Delta_r H(z) = 100 \times \left| H - H_{\Lambda \text{CDM}} \right| / H_{\Lambda \text{CDM}}$ as a function of the redshift $z$ for the two different initial conditions, as shown in the upper panel. The quantity $\Delta_r H(z)$ represents the relative difference between the results of our first interaction model and the $\Lambda$CDM model. From this lower panel, we observe that the largest difference with respect to the Hubble parameter $H_{\Lambda \text{CDM}}$ occurs for values of the redshift $z \lesssim 2$ in the case of the dot-dashed line, and at $z \simeq 0.5$ for the initial conditions defined by the solid line. Specifically, we observe that the relative difference $\Delta_r H \simeq 0$ occurs approximately at the present time for the initial variables defined by the solid line, at which point the Hubble rate $H$ coincides with $H_{\Lambda \text{CDM}}$ in this specific case.

Furthermore, the inset of the figure shows this plot, displaying the Hubble parameter $H(z)$ obtained from our interaction model alongside the Hubble parameter associated with the $\Lambda$CDM model ($H_{\Lambda \text{CDM}}(z)$, see Eq. (\ref{HE})), as a function of redshift. From this internal plot, we note that by comparing our interacting model with observational data, we can evaluate the agreement between the theoretical model and the empirical measurements of the Hubble parameter at various redshifts. The observational data corresponding to the $1 \sigma$ confidence intervals are provided in Appendix B, Table \ref{table:H(z)data}.

\begin{figure}[!tbp]
  \centering
    \includegraphics[width=0.6\linewidth]{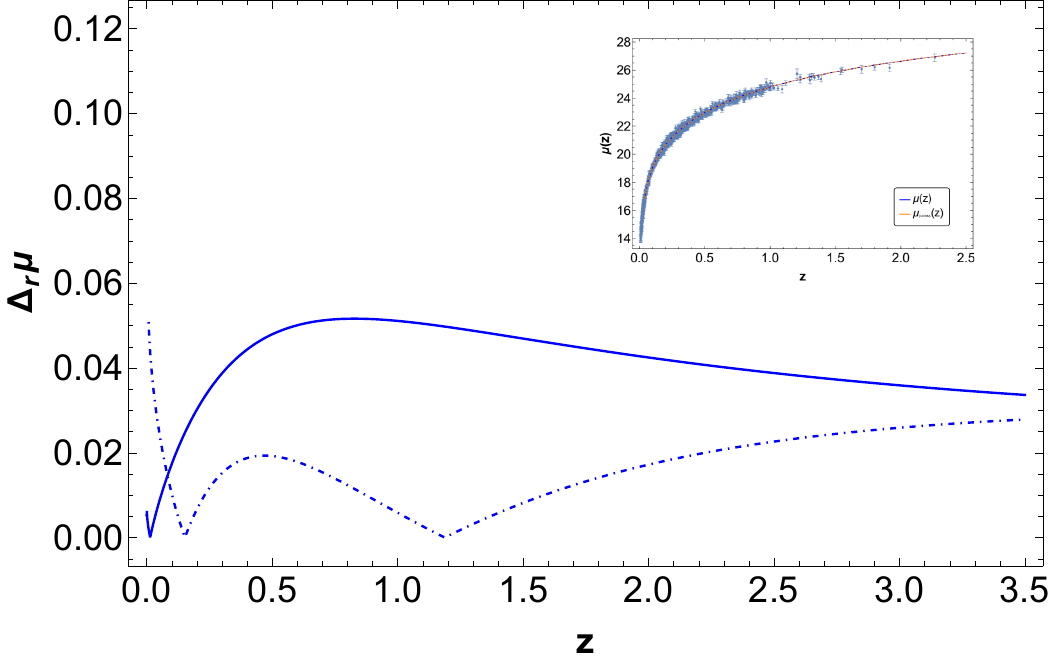} 
    \includegraphics[width=0.4\linewidth]{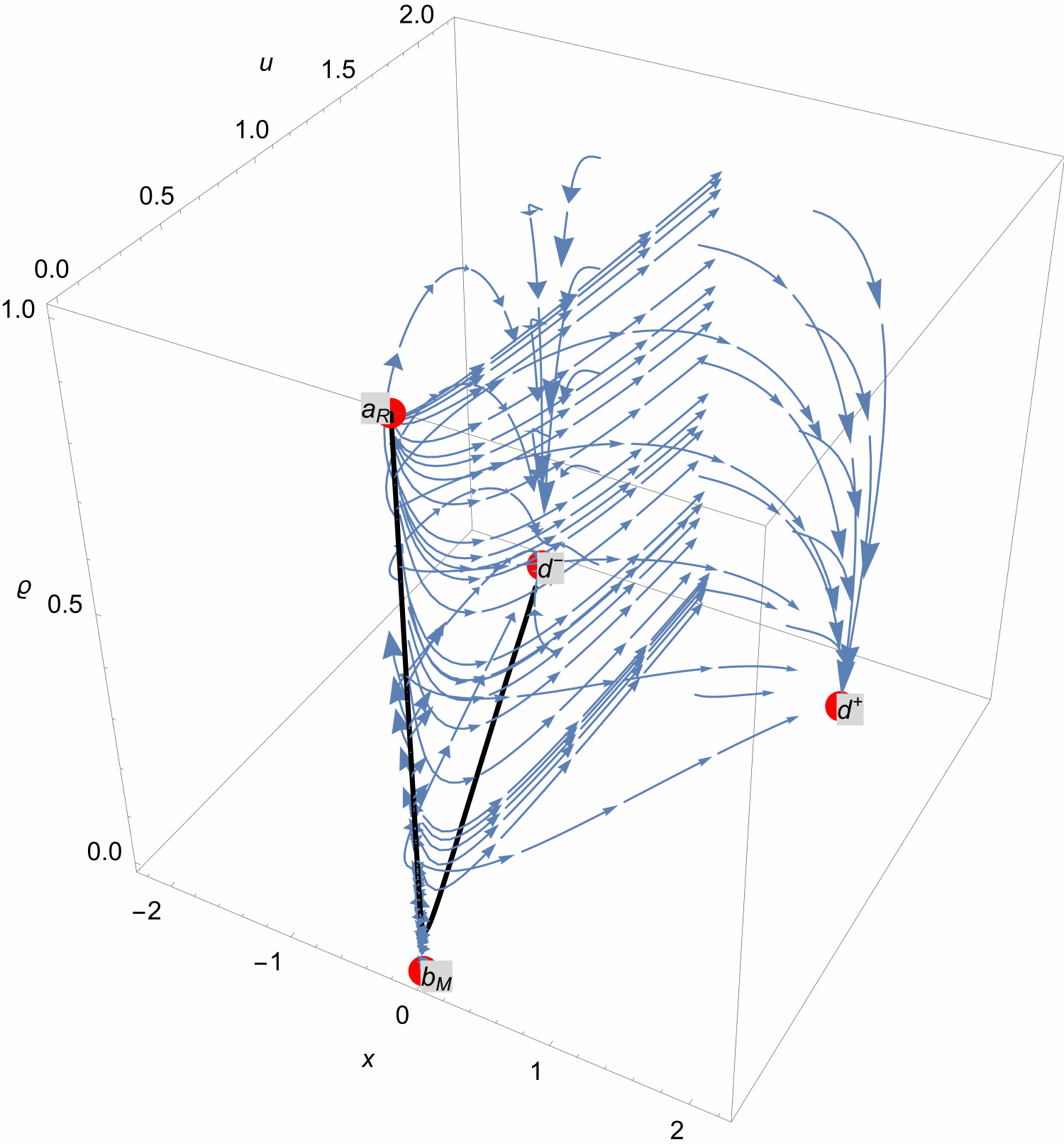} 
    \caption{\scriptsize{The upper panel shows the evolution of the relative difference $\Delta \mu_{r}$ with respect to the $\Lambda$CDM model as a function of the redshift $z$. Additionally, this panel presents the evolution of the distance modulus $\mu(z)$ for our first interacting model, alongside the distance modulus $\mu_{\Lambda CDM}$ for the $\Lambda$CDM model, both as functions of $z$. Two different sets of parameter values and initial conditions $x_i$, $y_i$, $u_i$, and $\varrho_i$ are considered, as in Fig. \ref{Fig1}.
The lower panel illustrates the evolution in phase space for our first interaction model $Q$, with parameters $\lambda = 10^{-3}$, $\sigma = 0.1$, $\alpha = -1$, and $\beta = 1.0 \times 10^{-3}$. Specifically, the black curve corresponds to the initial conditions $x_i = 1.0 \times 10^{-11}$, $y_i = 5.0 \times 10^{-13}$, $u_i = 1.25 \times 10^{-12}$, and $\varrho_i = 0.99983$.
    }
    } 
    \label{Fig2}
\end{figure}

On the other hand, the upper panel of Fig. \ref{Fig2} illustrates the evolution of the relative difference, $\Delta_r \mu$, in terms of the redshift for our first interaction model, $Q \propto \rho_m \dot{\phi}$, under different initial conditions (as used in Fig. \ref{Fig1}). The relative difference is defined with respect to the $\Lambda$CDM model, specifically as $\Delta_r \mu = 100 \times \left| \mu - \mu_{\Lambda \text{CDM}} \right| / \mu_{\Lambda \text{CDM}}$, where the distance modulus is given by Eq. (\ref{mu}). We note that the greatest difference occurs for values of $z \ll 2$. Additionally, within this upper panel, we show the evolution of the distance modulus $\mu$ for our first interaction model as a function of redshift. Here, we observe that our results are similar to those obtained in the $\Lambda$CDM model. We have contrasted these results with the latest Supernova Ia data \cite{Pan-STARRS1:2017jku}.

Additionally, the lower panel of Fig. \ref{Fig2} shows the evolution curves in the phase space for the specific case where the interaction parameter is $\beta = 1.0 \times 10^{-3}$, with the parameter values $\lambda = 10^{-3}$, $\sigma = 0.1$, $\alpha = -1$, and the initial conditions $x_i = 1.0 \times 10^{-11}$, $y_i = 5.0 \times 10^{-13}$, $u_i = 1.25 \times 10^{-12}$, and $\varrho_i = 0.99983$. In this panel, the phase space stream flow illustrates the trajectories $a_R \to b_M \to d^-$, demonstrating the evolution of the system. From the stability analysis of the critical points in the autonomous system, which includes the first interaction term, it is clear that the system evolves toward the attractors $d^\pm$. This behavior is clearly illustrated in the lower panel of Fig. \ref{Fig2}, where the trajectories converge to the attractor. The stability conditions confirm that these critical points represent stable solutions, corresponding to a dark energy-dominated era. This plot further demonstrates how the first interaction model naturally drives the cosmic dynamics toward a late-time attractor state, providing a consistent description of the universe's accelerated expansion.

\begin{figure}[!tbp]
  \centering
    \includegraphics[width=0.5\linewidth]{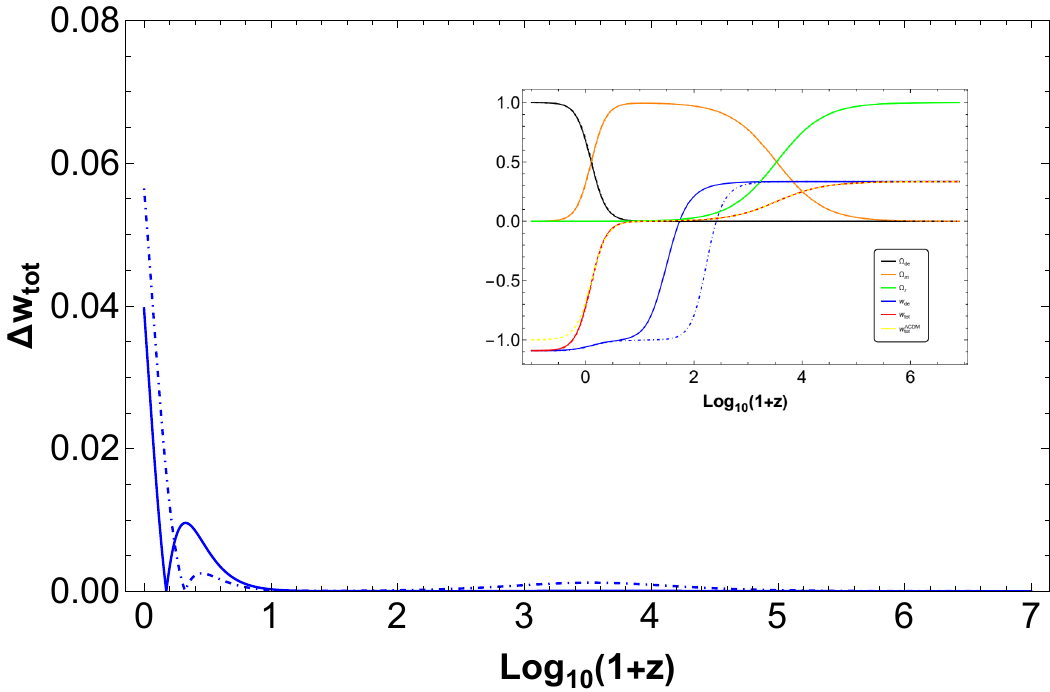} 
    \includegraphics[width=0.5\linewidth]{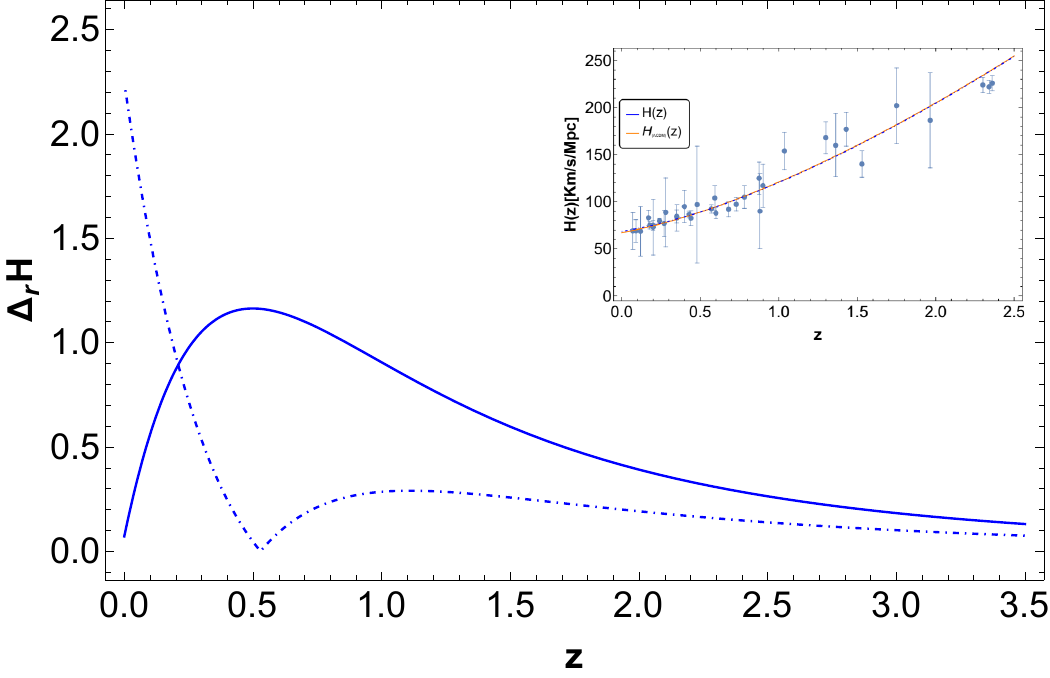} 
    \caption{\scriptsize{As in the previous case, the upper panel shows the evolution of the relative difference $\Delta w_{tot}$ as a function of Log${10}(1+z)$. Additionally, within this panel, we present the evolution of the different density parameters along with the EoS parameters as a function of Log${10}(1+z)$.
The lower panel illustrates the development of the relative difference $\Delta_r H$ associated with the Hubble parameter as a function of redshift $z$. The inset displays this plot, comparing the Hubble parameter $H(z)$ from our second interaction model with the Hubble parameter from the $\Lambda$CDM model ($H_{\Lambda CDM}(z)$, see Eq.(\ref{HE})), as a function of redshift, along with the observational data.
In both plots, we have used two different sets of initial conditions for the variables $x_i$, $y_i$, $u_i$, and $\varrho_i$, represented by the dot-dashed and solid lines, respectively.
    }
    } 
    \label{Fig3}
\end{figure}

\begin{figure}[!tbp]
  \centering
    \includegraphics[width=0.6\linewidth]{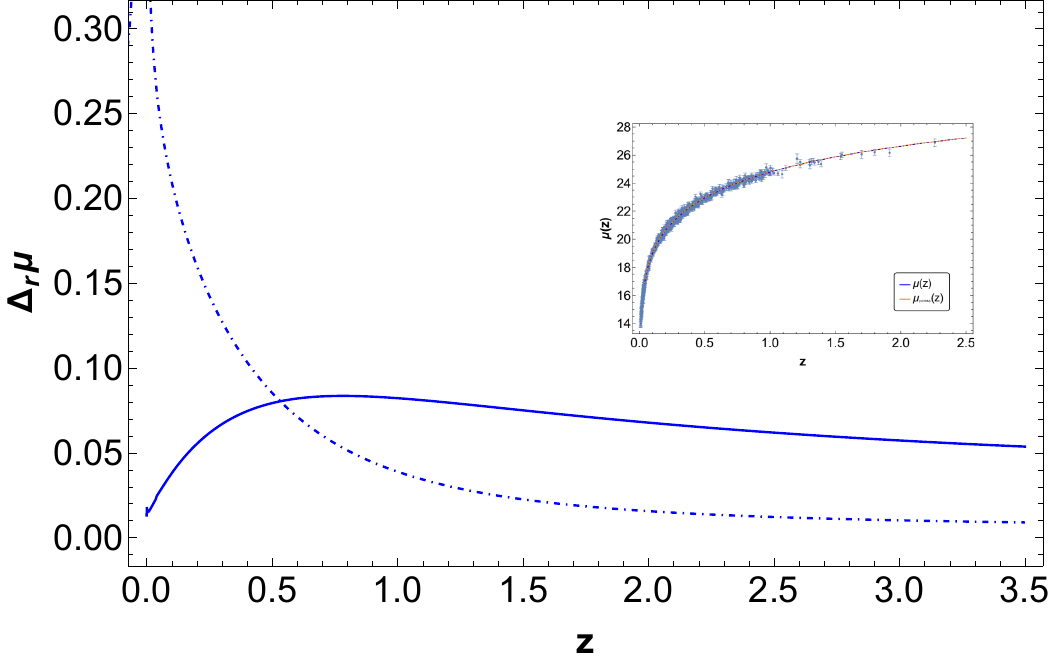} 
    \includegraphics[width=0.4\linewidth]{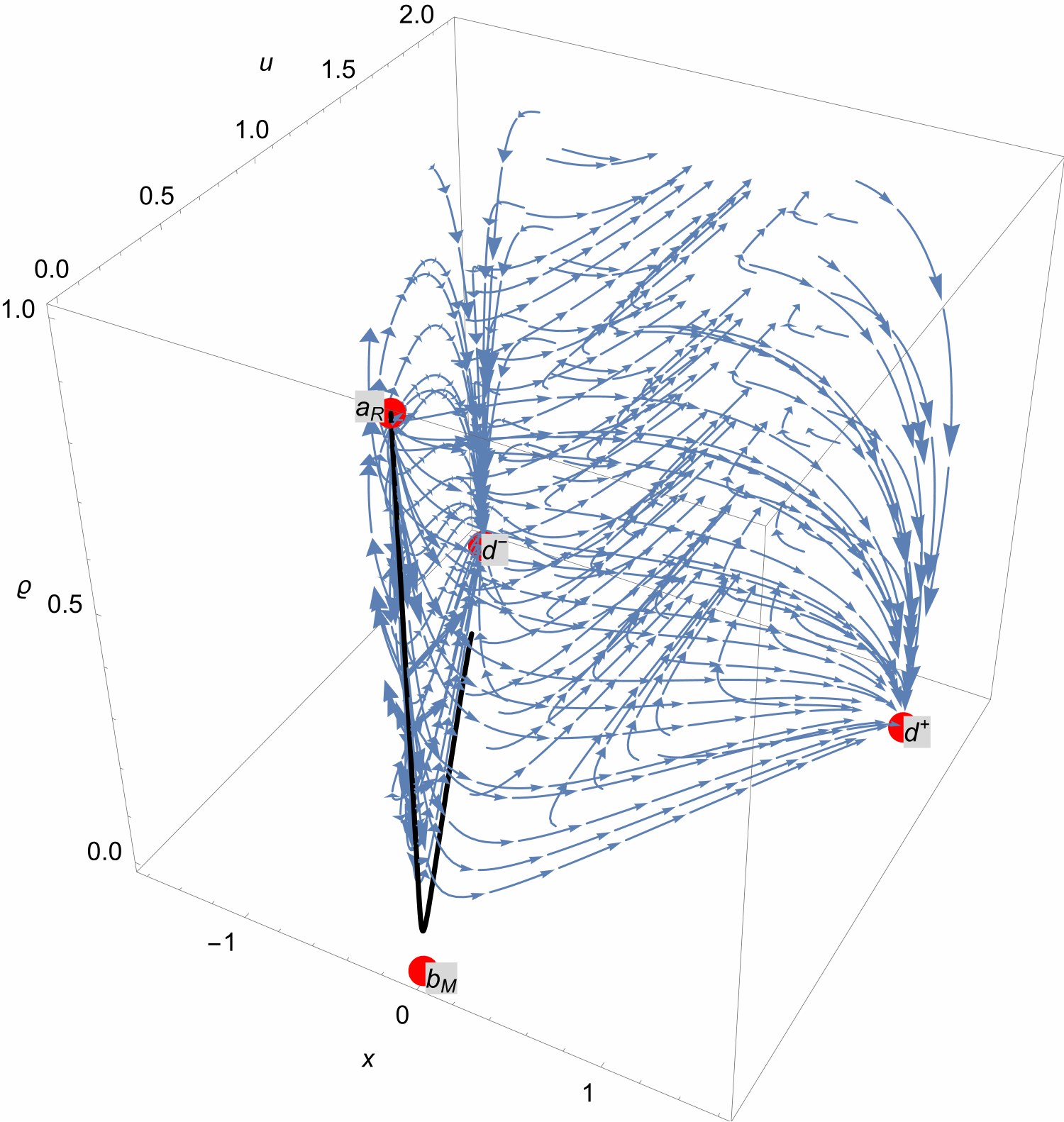} 
    \caption{\scriptsize{As in the previous case, the upper panel shows the evolution of the relative difference $\Delta \mu_{r}$ with respect to the $\Lambda$CDM model as a function of redshift $z$. Additionally, within this panel, we display the evolution of the distance modulus $\mu(z)$ for our second interacting model, where $Q \propto \rho_m H$.
The lower panel illustrates the evolution of the trajectories in the phase space, specifically the paths $a_R \to b_M \to d^-$ and the attractor $d^+$.
    }
    } 
    \label{Fig4}
\end{figure}

\begin{figure}[!tbp]
  \centering
    \includegraphics[width=0.5\linewidth]{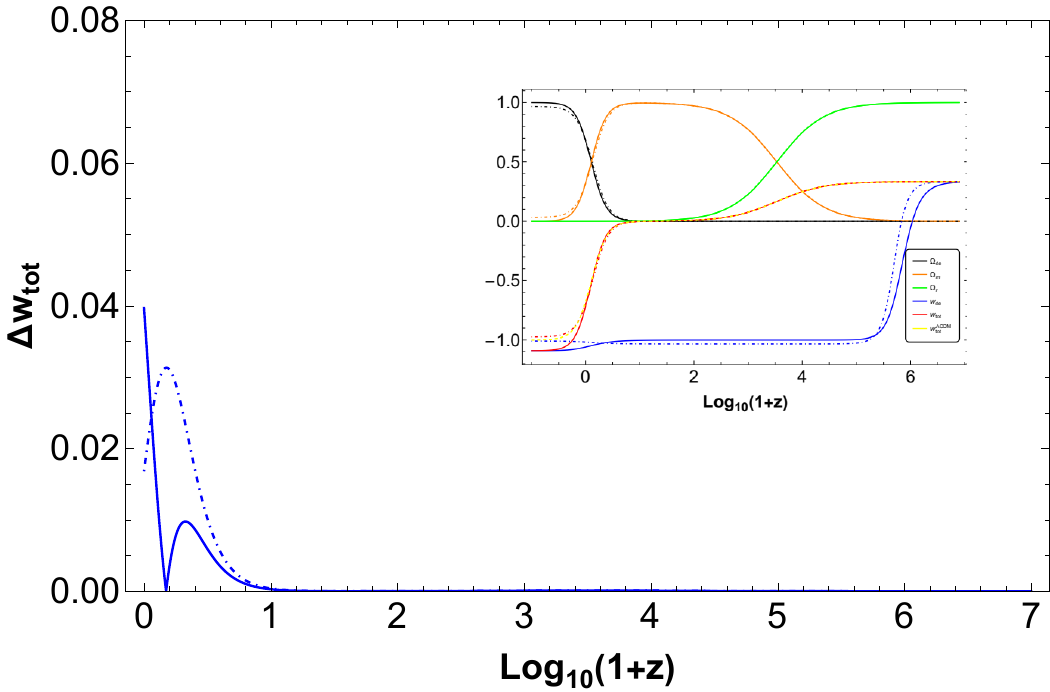} 
    \includegraphics[width=0.5\linewidth]{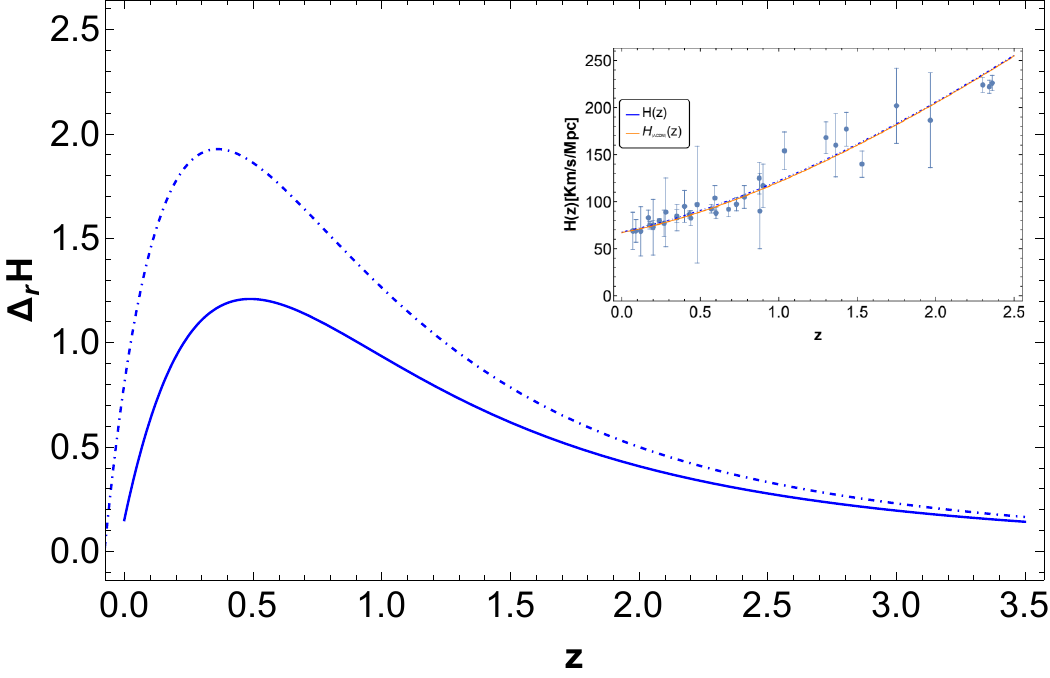} 
    \caption{\scriptsize{The upper panel displays the evolution of the relative difference $\Delta w_{tot}$ as a function of Log${10}(1+z)$. Additionally, within this panel, we show the evolution of the various density parameters along with the EoS parameters, all expressed in terms of the redshift function Log${10}(1+z)$.
The lower panel presents the development of the relative difference $\Delta_r H$ as a function of the redshift $z$. The inset in this panel shows the Hubble parameter $H(z)$ obtained from our third interaction model.
In both plots, we have used two values of the parameter $\beta$ and two different sets of initial conditions for the variables $x_i$, $y_i$, $u_i$, and $\varrho_i$, represented by dot-dashed and solid lines, respectively. Additionally, in both plots, we have fixed the values $\lambda = 0.1$, $\sigma = 0.1$, and $\alpha = -1$.
    }
    } 
    \label{Fig5}
\end{figure}

\begin{figure}[!tbp]
  \centering
    \includegraphics[width=0.6\linewidth]{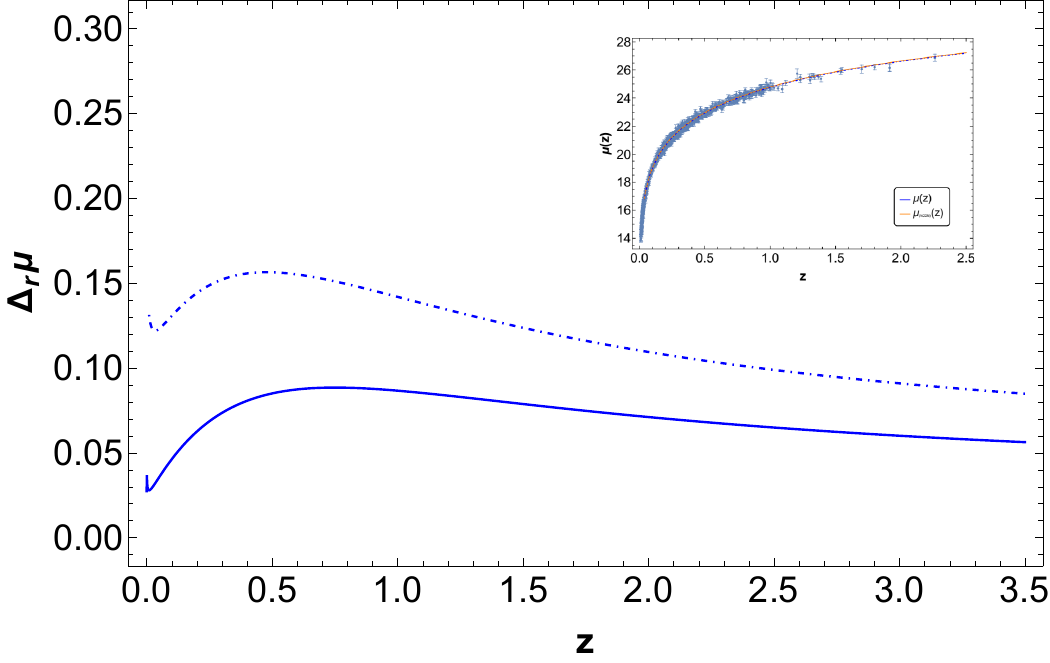} 
    \includegraphics[width=0.4\linewidth]{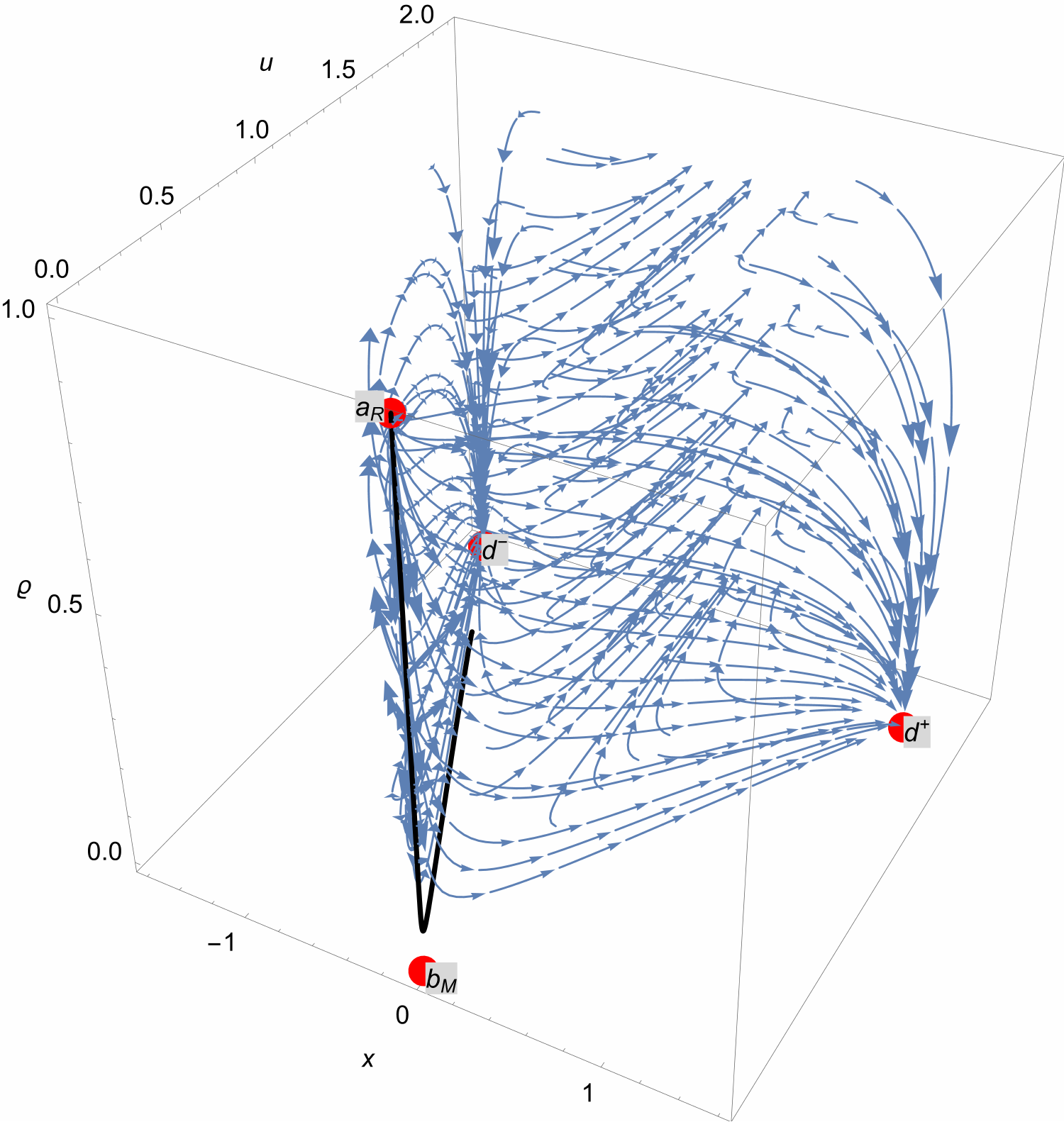} 
    \caption{\scriptsize{As in the previous cases, the upper panel shows the evolution of the relative difference $\Delta \mu_{r}$ with respect to $\Lambda CDM$ as a function of redshift $z$. Additionally, within this figure, we display the evolution of the distance modulus $\mu(z)$ for our interacting model, alongside the distance modulus for $\Lambda CDM$, denoted by $\mu_{\Lambda CDM}$, as a function of $z$.
The lower panel illustrates the evolution of the phase space trajectories and the attractor $d^+$ for our third interaction model, where $Q \propto \rho_{de} H$.
    }
    } 
    \label{Fig6}
\end{figure}

\subsubsection{Interaction $Q=3\beta\rho_m\,H$}

In this subsection, we present the numerical results for our second interaction, where the interaction term is given by $Q \propto \rho_m H$. As with the previous interaction, the upper panel of Fig.\ref{Fig3} illustrates the evolution of the absolute difference $\Delta w_{tot}$, which represents the deviation of the total EoS parameter compared to the $\Lambda$CDM model.

In this analysis, two distinct sets of parameter values and initial conditions were considered. The first set, represented by the dot-dashed line, corresponds to the values: $\lambda = 0.1$, $\sigma = 0.1$, $\alpha = -1$, $\beta = 1.8 \times 10^{-8}$, and the initial variables $x_i = 1.0 \times 10^{-11}$, $y_i = 5.0 \times 10^{-13}$, $u_i = 1.247 \times 10^{-12}$, and $\varrho_i = 0.99983$. The second set, represented by the solid line, is associated with the parameter values $\lambda = 0.1$, $\sigma = 0.1$, $\alpha = -1$, $\beta = -1.8 \times 10^{-5}$, and the initial variables $x_i = 1.0 \times 10^{-11}$, $y_i = 5.0 \times 10^{-13}$, $u_i = 1.19 \times 10^{-12}$, and $\varrho_i = 0.99983$.

From the upper panel of Fig.\ref{Fig3}, we observe that at the present time, the absolute difference $\Delta w_{tot}$ reaches its maximum value, indicating that the interaction model with $Q \propto \rho_m H$ exhibits the largest deviation from the $\Lambda$CDM model at this redshift. For values of Log${10}(1+z) \gtrsim 1$, the related difference $\Delta w{tot} \simeq 0$, suggesting that $w_{tot}$ from the second interaction closely resembles $w_{tot}^{\Lambda CDM}$.

Additionally, within this upper panel, we show the evolution of the density and EoS parameters as a function of Log${10}(1+z)$. From the internal plot, we observe that for the second interaction, $Q \propto \rho_m H$, the EoS parameter $w{tot}$ becomes negative for values of Log${10}(1+z) \lesssim 0.5$, and the accelerated phase occurs when this parameter drops below $-1/3$. Moreover, we note that the EoS parameter related to dark energy, $w{de}$, is negative for values of Log$_{10}(1+z) \lesssim 2.5$.

In particular, from this internal plot, we find that at the present time, the EoS parameter for dark energy is $w_{de} \simeq -0.8$, irrespective of the initial conditions (dot-dashed or solid line) associated with the autonomous system.

Furthermore, the lower panel of Fig.\ref{Fig3} presents the relative difference $\Delta_r H(z)$ for our second interaction model, where $Q \propto \rho_m H$, as a function of the redshift $z$ for the two different initial conditions shown in the upper panel. From this lower panel, we observe that the most significant difference with respect to the Hubble parameter $H_{\Lambda CDM}$ occurs at specific redshift values: $z \lesssim 2$ for the dot-dashed line and $0.1 \lesssim z \lesssim 1.5$ for the initial conditions associated with the solid line.

Additionally, we note that for the initial conditions defined by the dot-dashed line, the Hubble parameter at present (at $z = 0$) exhibits the largest deviation from the Hubble parameter of the $\Lambda$CDM model.

The inset further highlights this plot, showing the Hubble parameter $H(z)$ obtained from our second interaction model, alongside the Hubble parameter associated with the $\Lambda$CDM model ($H_{\Lambda CDM}(z)$, see Eq.(\ref{HE})) as a function of redshift $z$. From this internal plot, we observe that by comparing our interacting model with observational data, we can assess the agreement between the theoretical model and empirical measurements of the Hubble parameter for redshifts $z \gtrsim 2$ (as confirmed by $\Delta_r H$). The observational data are provided in confidence intervals of $1 \sigma$, as described in Appendix B, Table \ref{table:H(z)data}.

In addition, the upper panel of Fig.\ref{Fig4} shows the evolution of the relative difference $\Delta_r \mu$ as a function of redshift for our second interaction model, where $Q \propto \rho_m H$, with the same initial conditions considered in Fig.\ref{Fig3}. As before, this quantity is defined as the relative difference with respect to the $\Lambda CDM$ model, with the distance modulus given by Eq.(\ref{mu}).

We observe that the greatest difference in the distance modulus occurs for values of $z \ll 1.5$, indicating that the largest deviation from the $\Lambda CDM$ model happens at low redshift. Furthermore, within this upper panel, we also show the evolution of the distance modulus $\mu$ for our second interaction model as a function of redshift $z$. From this plot, we note that our results are similar to those obtained in the $\Lambda CDM$ model.

Furthermore, the lower panel of Fig.\ref{Fig4} presents the evolution curves in the phase space for the specific case where the interaction parameter is $\beta = 1.8 \times 10^{-8}$, and the parameter values are $\lambda = 0.1$, $\sigma = 0.1$, $\alpha = -1$, along with the different initial variables (as shown in Fig.\ref{Fig3}).

In this case, the lower panel of Fig.\ref{Fig4} illustrates the phase space stream flow for the trajectories $a_R \to b_M \to d^-$. From the stability analysis of the critical points in the autonomous system that includes the second interaction term $Q \propto \rho_m H$, we observe that the system evolves toward the attractors $d^\pm$.

This behavior is clearly shown in the lower panel of Fig.\ref{Fig4}, where the different trajectories converge to the attractor. The stability conditions confirm that these critical points correspond to stable solutions, associated with a dark energy-dominated era. The plot further demonstrates how the second interaction model influences the cosmic dynamics, guiding them toward a late-time attractor state, thereby ensuring a consistent description of the universe's accelerated expansion.

\subsubsection{Interaction $Q=3\beta\rho_{de}\,H$}

In this subsection, we present the numerical results obtained for our third interaction, given by $Q \propto \rho_{de} H$.

For this interaction term, as before, the upper panel of Fig.\ref{Fig5} shows the evolution of the absolute difference $\Delta w_{tot}$ associated with the total EoS parameter as a function of Log$_{10}(1+z)$. As in previous models, we use two different sets of values for the parameters and initial conditions related to $\lambda$, $\sigma$, $\alpha$, $\beta$, and the initial variables $x_i$, $y_i$, $u_i$, and $\varrho_i$.

The dot-dashed line corresponds to the parameters $\lambda = 0.1$, $\sigma = 0.1$, $\alpha = -1$, $\beta = 1.0 \times 10^{-1}$, and the initial variables $x_i = 1.0 \times 10^{-11}$, $y_i = 9.0 \times 10^{-13}$, $u_i = 4.1 \times 10^{-13}$, and $\varrho_i = 0.99983$. The solid line is associated with the parameters $\lambda = 0.1$, $\sigma = 0.1$, $\alpha = -1$, $\beta = 1.0 \times 10^{-3}$, and the initial variables $x_i = -1.0 \times 10^{-11}$, $y_i = 5.0 \times 10^{-13}$, $u_i = 1.21 \times 10^{-12}$, and $\varrho_i = 0.99983$.

From Fig.\ref{Fig5}, we observe that at present (i.e., at $z=0$), the relative difference is $\Delta w_{tot}(z=0) \sim 0.03$ for both sets of initial conditions. This indicates that the total EoS parameter $w_{tot}$ for the third interaction is very close to $w_{tot}^{\Lambda CDM}$ at the current time. Furthermore, the plot shows that the maximum value of the absolute difference $\Delta w_{tot}$ occurs at an approximate value of Log${10}(1+z) \simeq 0.3$ for the initial conditions associated with the dot-dashed line, and at present (i.e., $z=0$) for the initial conditions corresponding to the solid line. Furthermore, the figure displays the evolution of the different density parameters, along with the EoS parameters $w{de}$, $w_{tot}$, and $w_{tot}^{\Lambda CDM}$ as a function of Log$_{10}(1+z)$.

In particular, as before, the upper panel shows the evolution of the fractional energy densities of dark energy $\Omega_{de}$, dark matter $\Omega_m$, and radiation $\Omega_r$, along with the equation of state (EoS) of dark energy $w_{de}$, the total EoS parameter $w_{tot}$, and the EoS parameter of the $\Lambda$CDM model as a function of Log${10}(1+z)$. From this internal figure, we observe that the EoS parameter $w{tot}$ becomes negative for values of Log${10}(1+z) \lesssim 0.8$. Specifically, at present ($z=0$), the EoS parameter for dark energy, $w{de}$, takes the value $w_{de} \simeq -1.05$, regardless of the initial conditions (represented by the dashed line) in the autonomous system. Furthermore, we find that the total EoS parameter at $z=0$ is $w_{tot} \simeq -0.8$, a value that is very similar to the $\Lambda$CDM model. This is further corroborated by the plot of $\Delta w_{tot}$, where $\Delta w_{tot}(z=0) \simeq 0.02$ for the dashed blue line and $\Delta w_{tot}(z=0) \simeq 0.04$ for the solid blue line.

However, the lower panel of Fig.\ref{Fig5} shows the relative difference $\Delta_r H(z) = 100 \times \left| H - H_{\Lambda CDM} \right| / H_{\Lambda CDM}$ as a function of the redshift $z$ for the same initial conditions as in the upper panel. From this lower panel, we observe that the largest difference in the Hubble parameter $H$ compared to $H_{\Lambda CDM}$ occurs for $z \lesssim 2$ for both sets of initial conditions. Specifically, the maximum relative difference $\Delta_r H \simeq 0$ occurs around $z = 0.5$. At present, for the initial conditions associated with the solid blue line, we find that the Hubble rate $H$ is nearly identical to $H_{\Lambda CDM}$, resulting in $\Delta_r H(z=0) \simeq 0$. In contrast, for the initial conditions represented by the dashed blue line, the largest difference occurs at $z = 0$.

Additionally, the inset displays the plot that compares the Hubble parameter $H(z)$ from our third interaction model with the Hubble parameter from the $\Lambda$CDM model, as a function of the redshift. From this inset plot, we observe that by comparing our interacting model with observational data, we can assess the agreement between the theoretical model and the observational measurements of the Hubble parameter at various redshifts. As before, the observational data are shown with $1 \sigma$ confidence intervals (see Appendix B, Table \ref{table:H(z)data}).

On the other hand, the upper panel of Fig.\ref{Fig6} shows the evolution of the relative difference, $\Delta_r \mu$, as a function of redshift for our third interaction model $Q \propto \rho_{de} H$, with two different initial conditions (as in Fig.\ref{Fig5}). From this upper panel, we observe that the largest difference occurs for values of $z \lesssim 1.5$. Furthermore, within this upper panel, we also present the evolution of the distance modulus $\mu$ for our interaction model as a function of redshift. It is important to note that our results agree excellently with those obtained in the $\Lambda$CDM model, as reported in\cite{Pan-STARRS1:2017jku}.

In addition, the lower panel of Fig.\ref{Fig6} presents the evolution curves in the phase space for the specific case in which the interaction parameter is $\beta = 1.0 \times 10^{-3}$, with the parameter values $\lambda = 0.1$, $\sigma = 0.1$, $\alpha = -1$, and the initial variables $x_i$, $y_i$, $u_i$ and $\varrho_i$ defined above. This panel exhibits the phase-space stream flow for the trajectories $a_R \to b_M \to d^-$, and from the stability analysis of the critical points in the autonomous system that includes this interaction term, we observe that the system evolves towards the attractors $d^\pm$. The trajectories corresponding to these attractors are shown in the lower panel of Fig.\ref{Fig6}. The stability conditions confirm that these critical points represent stable solutions. The figure further demonstrates how this interaction naturally drives the cosmic dynamics toward a late-time attractor state, providing a consistent description of the universe's accelerated expansion.

\section{Conclusions}\label{conclusion_f}

In this article, we have studied  the cosmological evolution of a dilatonic ghost condensate field, associated with dark energy, which interacts with dark matter via a source term denoted by $Q$. To investigate various interaction models between the dilatonic field and dark matter, we have analyzed three interaction types that are widely discussed in the literature: $Q\propto \rho_m\dot{\phi}$, $Q\propto \rho_m,H$, and $Q\propto \rho_{de},H$.

For each interaction model, we have found the critical points by satisfying the conditions related to the dynamical system: $dx/dN=dy/dN=du/dN=d\varrho/dN=0$ (see, Eqs.(\ref{dinsyseq1})-(\ref{dinsyseq5})). These critical points for each interaction model are shown in their respective tables. Additionally, we have determined the cosmological parameters for these critical points and written them in a subsequent table for each interaction.

In relation to the critical points, we have found that the point $a_R$ corresponds to the critical point associated with a scaling radiation epoch. For this critical point, we have obtained the following values for the parameters: $\Omega_r = 1$, $w_{de} = 0$, and the total equation of state (EoS) parameter $w_{tot} = 1/3$. Additionally, we have found that this point does not depend on the values of $u_{c}$, and also that the point $a_R$ is independent of the parameter $\lambda$, which is associated with the effective potential.

Furthermore, we have determined that the critical point $b_M$ corresponds to a matter-dominated era, where $\Omega_m = 1$ and the parameters satisfy $w_{de} = w_{tot} = 0$.

In addition, we have found that the critical points $d^{\pm}$ represent a dark energy-dominated solution, both leading to a de Sitter solution in which the EoS parameters are $w_{de} = w_{tot} = -1$. At these points, the two values $d^{\pm}$ correspond to an accelerated expansion for all parameter values.

Moreover, we have determined that the critical point $c$, for the special case in which the parameter $\alpha$ is a function of the parameters $\beta$ and $\sigma$ (i.e., $\alpha = f(\beta, \sigma)$), represents a matter-dominated era, where the parameters satisfy $\Omega_m = 1$ and $\Omega_{de} = 0$, respectively.
In particular, for the first interaction model, we have found that for $\alpha=-1$, we have obtained the result that the matter-dominated era takes place for values of $\beta$ given by $\beta=(1/2)[-\sigma\pm\sqrt{\sigma^2+4}]$.

Additionally, we have analyzed the characteristics of our different interacting models in explaining the current accelerated expansion of the universe. In this context, we have compared the obtained results with the most recent observational data from $H(z)$ and supernovae Ia (SNe Ia) observations. For each interaction $Q$, we have determined the evolution of the absolute difference $\Delta w_{tot}$ associated with the total EoS parameters as a function of Log$_{10}(1+z)$, where $z$ is the redshift (upper panel of the figures). In this analysis, we considered two different sets of parameter values and initial conditions, specifically for the parameters $\lambda$, $\sigma$, $\alpha$, $\beta$, and the initial variables $x_i$, $y_i$, $u_i$, and $\varrho_i$.

In the lower panel of the same figures, we included the quantity $\Delta_r H(z)$, which represents the relative difference between the results of our different interaction models and the $\Lambda$CDM model. Specifically, we determined the relative difference
$\Delta_r H(z) = 100 \times \left| H - H_{\Lambda CDM} \right|/H_{\Lambda CDM}$ as a function of the redshift $z$ for the same two different initial conditions. We observed that the difference $\Delta_r H(z) < 0.5$, indicating a very small difference compared to the standard $\Lambda$CDM model (also shown in the inner plot).

On the other hand, we have included, for each interaction model, the evolution of the relative difference in the distance modulus, $\Delta_r \mu$, as a function of the redshift, under different initial conditions (see Eq. (\ref{mu})). In relation to this relative difference, we found that for the different interaction models, the greatest difference occurs for redshifts $z \ll 3$. We observed that the values of $\Delta_r \mu$ vary depending on the interaction model, with the largest difference occurring for the interaction $Q \propto \rho_m,H$, around $z \sim 0$ (see the upper panel of Fig. (\ref{Fig4})).
Additionally, we determined the evolution curves in the phase space for the different interactions. For this analysis, we used various parameters and initial conditions specific to the interaction model studied.

 For the phase space, we have graphed the flow of the phase space stream for the trajectories $a_R \to b_M \to d^-$. Additionally, we performed a stability analysis of the critical points in the autonomous system that includes the different interaction terms. From this analysis, we determined that the system evolves towards the attractor $d^+$. We also observed that these trajectories converge to this attractor.
Thus, the stability conditions confirm that these critical points represent stable solutions, which correspond to a dark energy-dominated era.

It is important to note that there are some properties in this interacting model that warrant further investigation. In particular, we did not address the other interactions present in the literature. Similarly, we did not develop a study on the formation of structures in our interaction models, as described in Refs.\cite{Herrera:2004dh,Herrera:2016uci,delCampo:2013hka}. We plan to revisit these points and explore them in greater depth in the near future.

\begin{acknowledgments}
M. Gonzalez-Espinoza acknowledges the financial support of FONDECYT de Postdoctorado, N° 3230801. 
\end{acknowledgments}


\bibliography{bio}

\begin{appendix}

\section{Critical points for  interaction $Q=3\beta\rho_{de}\,H$}
\label{AppB}

To simplify the expressions and highlight the recurring terms, we define the following quantities:

\begin{align}
    B_1 &\coloneqq -2 \sqrt{3} \alpha^3 (3 + \beta)^3 + 9 \sqrt{3} \alpha^2 (-63 - 9 \beta + \beta^2) \sigma^2 + 27 \sqrt{3} \alpha (15 + \beta) \sigma^4  \nonumber\\
    &\quad + 27 \left(-2 \sqrt{3} \sigma^6 + \sqrt{\alpha^2 \sigma^2 B_2}\right), \\
    B_2 &\coloneqq 4 \alpha^3 (3 + \beta)^3 (4 + \beta) - \alpha^2 (-207 + 582 \beta + 295 \beta^2 + 38 \beta^3 + \beta^4) \sigma^2  \nonumber\\
    &\quad + 6 \alpha (-1 + 32 \beta + 16 \beta^2 + \beta^3) \sigma^4 - 9 (1 + \beta)^2 \sigma^6, \\
    B_3 &\coloneqq \left(\frac{\alpha^2 (3 + \beta)^2 - 3 \alpha (15 + \beta) \sigma^2 + 9 \sigma^4}{B_1}\right)^{1/3}, \\
    B_4 &\coloneqq 3 \sqrt{6} \alpha + \sqrt{6} \alpha \beta + 3 \sqrt{6} \sigma^2.
\end{align}

With the previous definitions, the main expressions of critical points are presented as follows:

\begin{eqnarray}
    x_{0}^{\pm} &=& \frac{1}{18 \alpha \sigma^2} \Bigg( 
    B_4 - 2^{5/6} 3^{2/3} B_3 - 2^{1/6} \big(
    -6 \sqrt{3} \alpha^3 (3 + \beta)^3 + 27 \sqrt{3} \alpha^2 (-63 - 9 \beta + \beta^2) \sigma^2 \nonumber \\
    && + 81 \sqrt{3} \alpha (15 + \beta) \sigma^4 
    + 81 \left(-2 \sqrt{3} \sigma^6 + \sqrt{\alpha^2 \sigma^2 B_2}\right)
    \big)^{1/3}
    \Bigg),
\end{eqnarray}
and
\begin{eqnarray}
    u_{0}^{\pm} = \frac{1}{2^{1/4} 3^{3/4} \sqrt{\alpha \left(\alpha (6 + \beta)^2 - 2 (3 + \beta) \sigma^2\right)}} 
    \Bigg( \sqrt{6} \Bigg(
    216 + 90 \beta + 9 \beta^2 - \frac{30 (\alpha (3 + \beta) + 3 \sigma^2)}{\alpha}  \nonumber\\
    - \frac{10 \beta (\alpha (3 + \beta) + 3 \sigma^2)}{\alpha} 
    - \frac{\beta^2 (\alpha (3 + \beta) + 3 \sigma^2)}{\alpha} + \frac{30 \cdot 2^{5/6} \cdot 3^{2/3} (\alpha^2 (3 + \beta)^2 - 3 \alpha (15 + \beta) \sigma^2 + 9 \sigma^4)}{B_1} \Bigg) + \nonumber\\
     \frac{10 \cdot 2^{5/6} \cdot 3^{2/3} \beta (\alpha^2 (3 + \beta)^2 - 3 \alpha (15 + \beta) \sigma^2 + 9 \sigma^4)}{B_1} + \frac{2^{5/6}  3^{2/3} \beta^2 (\alpha^2 (3 + \beta)^2 - 3 \alpha (15 + \beta) \sigma^2 + 9 \sigma^4)}{B_1} \Bigg). 
\end{eqnarray}

\section{Hubble's parameter data}\label{appen_B}

In this appendix, we present Hubble's parameter data for the  redshift range $0.01 < z < 2.37$:
\begin{table}[!b]
\caption{redshift versus Hubble's parameter, including references.  }
\label{table:H(z)data}
\renewcommand{\tabcolsep}{0.7pc} 
\renewcommand{\arraystretch}{0.7} 
\begin{tabular}{@{}lllll}
\hline \hline
  $\;\; z$    &  $ H(z) \;$ ($\frac{km/s}{\text{Mpc}}$ ) &  Ref. \\
\hline
$0.07$      & $ \; \qquad 69     \pm 19.6 $      &   \cite{zhang2014} \\
$0.09$      & $ \; \qquad 69     \pm 12 $      & \cite{simon2005} \\
$0.100$     & $ \; \qquad 69     \pm 12 $      & \cite{simon2005} \\
$0.120$     & $ \; \qquad 68.6     \pm 26.2$       & \cite{zhang2014} \\
$0.170$     & $ \; \qquad 83     \pm 8$       & \cite{simon2005} \\
$0.179$     & $ \; \qquad 75     \pm 4$       & \cite{moresco2012} \\
$0.199$     & $ \; \qquad 75     \pm 5$        & \cite{moresco2012} \\
$0.200$     & $ \; \qquad 72.9     \pm 29.6$        &  \cite{zhang2014} \\
$0.270$     & $ \; \qquad 77     \pm 14$      & \cite{simon2005} \\
$0.280$     & $ \; \qquad 88.8     \pm 36.6$      & \cite{zhang2014} \\
$0.320$     & $ \; \qquad 79.2   \pm 5.6$     & \cite{cuesta2016}\\
$0.352$     & $ \; \qquad 83     \pm 14$      & \cite{moresco2012} \\
$0.3802$    & $ \; \qquad 83     \pm 13.5$      & \cite{moresco2012} \\
$0.400$     & $ \; \qquad 95     \pm 17$      & \cite{simon2005} \\
$0.4004$    & $ \; \qquad 77     \pm 10.2$      & \cite{moresco2012} \\
$0.4247$    & $ \; \qquad 87.1     \pm 11.2$      & \cite{moresco2012} \\
$0.440$     & $ \; \qquad 82.6   \pm 7.8$     & \cite{blake2012} \\
$0.4497$    & $ \; \qquad 92.8   \pm 12.9$     & \cite{moresco2012} \\
$0.470$     & $ \; \qquad 89   \pm 50$     & \cite{ratsim} \\
$0.4783$    & $ \; \qquad 80.9   \pm 9$     & \cite{moresco2012} \\
$0.480$     & $ \; \qquad 97     \pm 62$      & \cite{stern2010} \\
$0.570$     & $ \; \qquad 100.3  \pm 3.7$     & \cite{cuesta2016} \\
$0.593$     & $ \; \qquad  104   \pm 13$      & \cite{moresco2012} \\
$0.600$     & $ \; \qquad 87.9   \pm 6.1$     & \cite{blake2012} \\
$0.680$     & $ \; \qquad 92     \pm 8$       & \cite{moresco2012} \\
$0.730$     & $ \; \qquad 97.3   \pm 7 $      & \cite{blake2012} \\
$0.781$     & $ \; \qquad 105    \pm 12$      & \cite{moresco2012} \\
$0.875$     & $ \; \qquad 125    \pm 17$      & \cite{moresco2012} \\
$0.880$     & $ \; \qquad 90     \pm 40$      & \cite{stern2010} \\
$0.900$     & $ \; \qquad 117    \pm 23$      & \cite{simon2005} \\
$1.037$     & $ \; \qquad 154    \pm 20 $     & \cite{moresco2012} \\
$1.300$     & $ \; \qquad 168    \pm 17 $     & \cite{simon2005} \\
$1.363$     & $ \; \qquad 160    \pm 33.6$    & \cite{moresco2015}\\
$1.430$     & $ \; \qquad 177    \pm 18$      & \cite{simon2005}\\
$1.530$     & $ \; \qquad 140    \pm 14$      & \cite{simon2005}\\
$1.750$     & $ \; \qquad 202    \pm 40$      & \cite{simon2005}\\
$1.965$     & $ \; \qquad 186.5  \pm 50.4$    & \cite{moresco2015}\\
$2.340$     & $ \; \qquad 222    \pm 7 $      & \cite{delubac2014}\\
$2.360$     & $ \; \qquad 226    \pm  8$      & \cite{font-ribera2014}\\
\hline
\end{tabular}\\
 \end{table}

\end{appendix}

\end{document}